\documentclass[preprint,12pt]{elsarticle}

\usepackage{amssymb}

\usepackage{lineno}
\usepackage{amsmath}
\usepackage{comment}
\usepackage{textcomp}
\journal{Nuclear Instruments Methods Ser. B}

\usepackage{subcaption}

\usepackage{color, soul}

\begin{document}

\begin{frontmatter}



\title{Parameterization of dose profiles of therapeutic minibeams of protons, $^{4}$He, $^{12}$C, and $^{16}$O \tnoteref{label1}}

\author[INR]{Savva~Savenkov}
\author[INR]{Alexandr~Svetlichnyi}
\author[INR]{Igor~Pshenichnov\corref{correspondingauthor}}

\cortext[correspondingauthor]{Corresponding author}
\ead{pshenich@inr.ru}

\affiliation[INR]{organization={Institute for Nuclear Research of the Russian Academy of Sciences},
addressline={60-letia Okatyabrya, 7a},
city={Moscow},
postcode={117312},
state={},
country={Russia}}


\begin{abstract}
Single minibeams of protons, $^{4}$He, $^{12}$C and $^{16}$O in water were modeled with Geant4, and their dose distributions were parameterized with double-Gauss-Rutherford (DGR) functions. Dose distributions from arrays of 16 parallel minibeams centered on a rectangular or hexagonal grid were constructed from the parameterized minibeam profiles to simulate the lateral convergence of the minibeams resulting in a homogeneous dose field in the target tumor volume. Peak-to-valley dose ratios (PVDR) and dose-volume histograms (DVH) were calculated for the parameterized dose distributions and compared with those obtained directly from Geant4 modeling of minibeam arrays. The similarity of the results obtained by these two methods suggests that the fast calculation of dose profiles of minibeam arrays based on the DGR parameterizations proposed in this work can replace the time-consuming MC modeling in future preclinical studies and also in the development of treatment planning systems for minibeam therapy.
\end{abstract}




\end{frontmatter}


\section{Introduction}\label{sec:intro}

Cancer particle therapy using proton or heavy ion beams is  considered as an advanced treatment option compared to X-ray therapy. Because of the specific depth-dose distribution known as the spread-out Bragg peak, therapy with these heavy charged particles provides reduced doses at the entrance to the patient's body and to the healthy tissues surrounding the tumor~\cite{Mohan2022}. At more than 100 proton therapy centers operating worldwide~\cite{Yan2023}, thorough treatment planning procedures ensure high tumor dose conformality while sparing healthy tissues and critical organs. Nevertheless, there is a general consensus that there is a need for continued improvement in the state of the art of proton therapy~\cite{Mohan2022}. The creation of internal structures of time and spatial dose distributions, known as FLASH and minibeam therapies, respectively, are both considered promising for advanced treatment with protons with reduced complications in healthy tissues~\cite{Mazal2020}. 

Very high dose rates of 9.3~Gy/s or 930~Gy/s~\cite{Rudigkeit2024} are typically considered for FLASH therapy to enhance biological and clinical effects. Minibeam proton therapy~\cite{Prezado2013} relies on the use of arrays of thin parallel proton minibeams of 0.3--1.1~mm full width at half maximum (FWHM) with center-to-center spacing of 1--3.5~mm~\cite{Zlobinskaya2013,Meyer2019}. In Ref.~\cite{Sammer2017}, larger center-to-center distances up to 6.84~mm were considered for arrays of unidirectional pencil minibeams of protons. By using an innovative geometry of interlaced minibeams from multiple directions, the center-to-center distance between pencil minibeams can be increased up to 8.9~mm~\cite{Sammer2021a}. Since many cells of normal tissue between individual minibeams receive a minimal dose, this facilitates the recovery of the tissue~\cite{Zlobinskaya2013,Girst2016,Prezado2017b} compared to its irradiation with a homogeneous entrance dose field. A combined temporal and spatial fractionation of therapeutic proton fields was also studied recently~\cite{Sammer2021,Sammer2022}. 

As shown in the evaluation of the first treatment plan generated by Monte Carlo modeling for proton minibeam therapy of a virtual patient with either glioma or meningioma~\cite{Lansonneur2020}, planar minibeams converge laterally at greater depths, providing a homogeneous dose field across the target tumor volume as in conventional proton therapy. Spatially fractionated dose fields from circular minibeams of $^{4}$He~\cite{Schneider2019a}, planar minibeams of $^{12}$C and $^{16}$O~\cite{Gonzalez2017}, as well as from planar and grid minibeams of $^{12}$C, $^{16}$O, $^{20}$Ne, $^{28}$Si, $^{40}$Ar, $^{56}$Fe~\cite{Gonzalez2018} have also been modeled for possible applications in heavy-ion minibeam therapy. 

Monte Carlo (MC) calculations of dose distributions for arrays of mini\-beams are very demanding in terms of the CPU time due to  large dose gradients on the sub-millimeter scale. This requires a sub-millimeter mesh size of the dose scoring grid to ensure accurate calculations of the peak-to-valley dose ratio (PVDR)~\cite{Prezado2013} considered as one of the main metrics of spatially fractionated dose fields delivered by arrays of minibeams. For example, a bin or voxel size of 0.025~mm may be suggested for reliable calculations of the lateral dose distribution from a single circular minibeam of 0.3--0.5~mm FWHM. It is therefore important that simulation settings such as the cut in range, which defines the minimum energy of the secondary particles in the modeling~\cite{Grevillot2010}, and the track step length are sufficiently short for accurate calculations of the energies deposited by primary and secondary particles in each of such small voxels. This requires multiple calculations of the ionization energy loss of particles for each $\sim 0.025$~mm of their track length, as well as the generation of enormous numbers of low-energy delta-electrons produced by charged particles in simulation.

As known, modern treatment planning systems in particle therapy are based on beam data libraries that typically consist of dose profiles of pencil-like beams of 4--9~mm FWHM ~\cite{Grevillot2011}. For example, the Gauss-Rutherford functions with four parameters have been found to be  suitable~\cite{Bellinzona2015} to approximate the measured and calculated lateral dose profiles of proton pencil-like beams. These pre-calculated dose profiles are then used to compose patient-specific dose fields with treatment planning systems~\cite{Parodi2012}. In real treatments, the required dose fields are delivered by active beam scanning systems as a sequence of pencil-like beams of different energies directed from certain positions and angles strictly following a prescribed treatment plan.

The Gaussian and double-Gaussian approximations were used in Refs.~\cite{Sammer2017} and~\cite{Sammer2021a}, respectively, to describe the radial dependence of the dose profiles of individual minibeams delivered to  tumors, and the geometry of irradiation with minibeam arrays has been optimized to achieve a better biological outcome. Following~\cite{Sammer2021a}, in the present work the double-Gaussian approximations were first applied to proton minibeams and then supplemented with an additional term to improve the dose description far from the beam axis. In addition to proton minibeams, the dose profiles of minibeams of $^4$He, $^{12}$C and $^{16}$O have been approximated in the present work.  

With the replacement of traditional pencil-like beams by arrays of mini\-beams in preclinical studies or treatment planning, it will be necessary to supplement the dose distributions stored in libraries with the functions describing the depth dependence of the PVDR. This will make it possible to quantify the biological impact of minibeam arrays on tissues as a function of the geometrical arrangement of the minibeams, their width, and the center-to-center distance.  Currently, the systematic evaluation of the influence of minibeam width and spacing on the response to irradiation is missing for most of the organs and tumor tissues~\cite{Mazal2020,Prezado2021}. The studies of the dependence of the skin reactions on minibeam size of single minibeams and minibeam arrays presented in Refs.~\cite{Sammer2019,Sammer2019a} can be mentioned as important steps towards such a systematic evaluation.

A search for optimal geometric arrangements of minibeams within mini\-beam arrays in preclinical studies or treatment planning requires multiple runs of MC modeling of each specific configuration. Instead, a library of dose profiles and corresponding PVDRs can be collected by (1) MC modeling dose profiles of individual minibeams; (2) approximating these profiles with relevant functions; (3) fast constructing doses from minibeam arrays and corresponding PVDRs based on the approximations of individual minibeams. 

This work aims at implementing steps (1)--(3) for possible future applications in minibeam therapy with protons, $^{4}$He, $^{12}$C and $^{16}$O. In Sec.~\ref{sec:validation}, the propagation of pencil-like beams of protons and $^{12}$C in water has been modeled with the Geant4 toolkit. The calculated dose profiles were compared with measurements to validate our application based on this toolkit. In Sec.~\ref{sec:approxmating_functions} double-Gauss (DG) and double-Gauss-Rutherford (DGR) functions were introduced to approximate the lateral profiles of single circular minibeams (or pencil minibeams, following~\cite{Sammer2017}) in a water phantom calculated by Monte Carlo modeling with Geant4 as described in Sec.~\ref{sec:single_minibeams}. In the same  section, the approximation accuracy was evaluated by considering the relative difference between the minibeam dose distribution from MC and its approximating function at all locations in the water phantom. In Sec.~\ref{sec:minibeam_arrangements} the dose fields of arrays of 16 parallel circular minibeams of protons, $^{4}$He, $^{12}$C and $^{16}$O propagating in water were obtained by MC modeling and also independently constructed from the approximated dose profiles of the corresponding individual minibeams. 
In Sec.~\ref{sec:PVDR16}, the values of PVDR as a function of depth in water were calculated for the dose fields from minibeam arrays by direct MC modeling and also by superimposing the approximations obtained for individual minibeams. In Sec.~\ref{sec:DVH16}, the dose-volume histograms (DVHs) were calculated and compared for arrays of minibeams at the entrance of the phantom and in the vicinity of the Bragg peak. Finally, a summary of the obtained results is given in Sec.~\ref{sec:conc}, and the possibility to use the fast calculations based on the DGR
approximations instead of the time-consuming Monte-Carlo modeling is underlined.

\section{Geant4 calculation of dose in water from pencil-like beams of protons and $^{12}$C}\label{sec:validation}

In the present work the propagation of pencil-like beams of protons and $^{12}$C in water was modeled with Geant4 toolkit~\cite{Agostinelli2003,Allison2006,Allison2016} of version 10.3. The physics lists used in the modeling were the same as in Ref.~\cite{Dewey2017}.  Electromagnetic processes were modeled with {EMStandard\_opt3} list, Binary Cascade (BIC) model was involved for proton- and neutron-induced nuclear reactions and Quantum Molecular Dynamics (QMD) model was used in simulating  nucleus-nucleus collisions. It was shown~\cite{Sato2022} that Geant4 QMD model describes well the measured distributions of secondary fragments in depth, angle and energy  from the interactions of 400A~MeV $^{12}$C in water. 

Our Geant4-based application has been additionally validated by comparison of calculated lateral profiles with the dose profiles of pencil-like beams of 182.66~MeV protons and 350.84A~MeV $^{12}$C in water 
measured in Ref.~\cite{Schwaab2011}. The profiles were measured at the depth of the plateau region (35.5~mm) and peak (220.5~mm) for the proton beam, and at the depth of the peak (210.2~mm) and tail region (260.2~mm) for the beam of $^{12}$C.
 
For the purpose of the comparison with the data, the 3D dose distributions in the Cartesian coordinates for proton beam of 9.85~mm FWHM and $^{12}$C beam of 3.82~mm FWHM were calculated in rectangular water phantoms with dimensions of $60\times 60 \times 250$~mm$^3$ and  $60\times 60 \times 400$~mm$^3$, respectively, to reproduce the measurement conditions~\cite{Schwaab2011}.  The X-axis was directed along the beam, and the voxel size of $1\times 1 \times 1$~mm$^3$ was implemented in Geant4 modeling to score the dose distributions in the phantoms. The values of of initial FWHM of the beams in calculations were estimated from the dispersion $\sigma_1$ of the lateral beam profiles obtained from the fit of data at the entrance to the phantom~\cite{Schwaab2011}. The Gaussian lateral beam profiles were assumed in the modeling. The slices of 3D dose distributions of 5~mm thick at the plateau and tail, but 1~mm thick at the peak,  were projected on Z-axis to obtain the lateral beam profiles at given depth. Such a selection of the slice thickness is motivated by a characteristic diameter of an active volume of 30~mm$^3$ of each of the ionization chambers assembled in arrays, as well as by the step size in measurements~\cite{Schwaab2011}. 

The intrinsic angular beam divergence was disregarded in the calculations with pencil-like beams, since, according to Ref.~\cite{Grevillot2010}, the effect of multiple Coulomb scattering on lateral dose spreading in water is expected to be more significant than the divergence of the initial beam delivered by an accelerator. The calculated depth of the Bragg peak (220.65~mm) for protons was found in a perfect agreement with the measured one (220.5~mm). However, the measured (210.2~mm) and calculated (222.5~mm) positions of the Bragg peak for $^{12}$C diverged by 12.3~mm. One can attribute such a shift to the water-equivalent thickness of a ripple filter placed in front of the water phantom during the measurements with $^{12}$C. Since the water-equivalent thickness of the filter was not reported in Ref.~\cite{Schwaab2011}, the lateral profile for the peak was calculated at its depth obtained in modeling (222.5~mm). 
\begin{figure}[htb!]
    \centering
    \begin{minipage}{0.49\linewidth}
    \includegraphics[width = 1.15\linewidth]{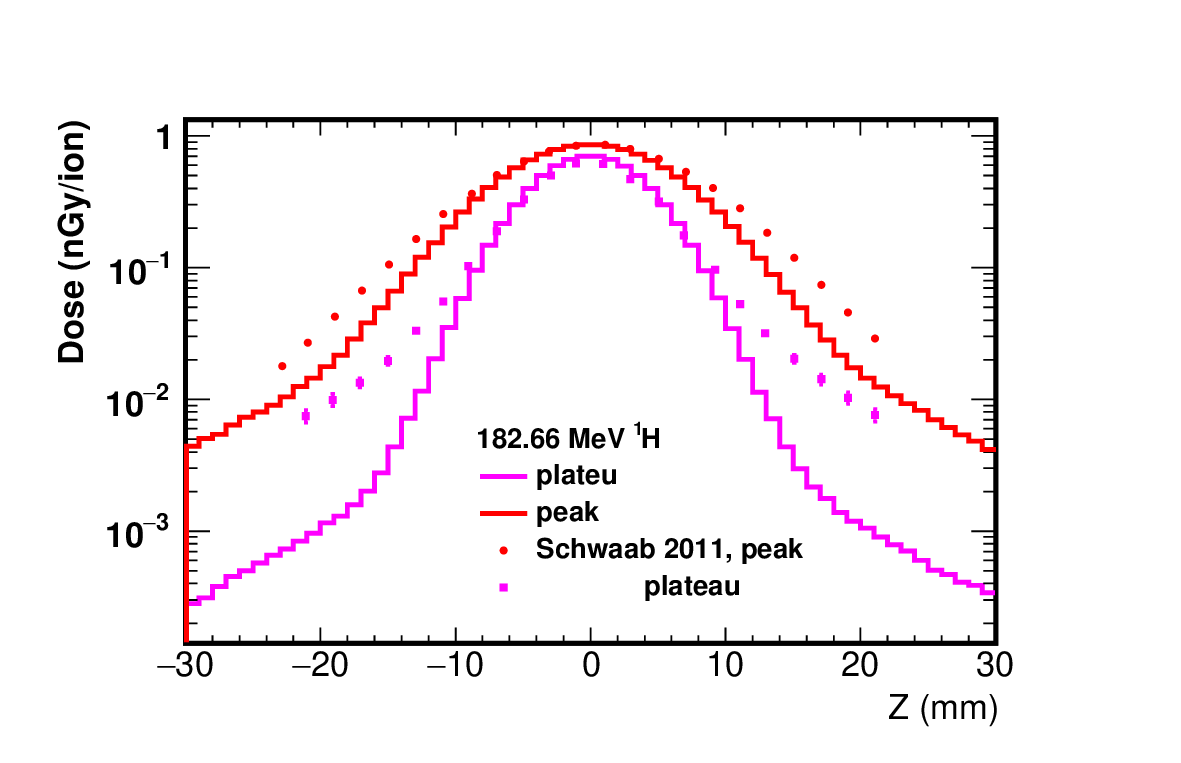}
    \end{minipage}
    \begin{minipage}{0.49\linewidth}
    \includegraphics[width = 1.15\linewidth]{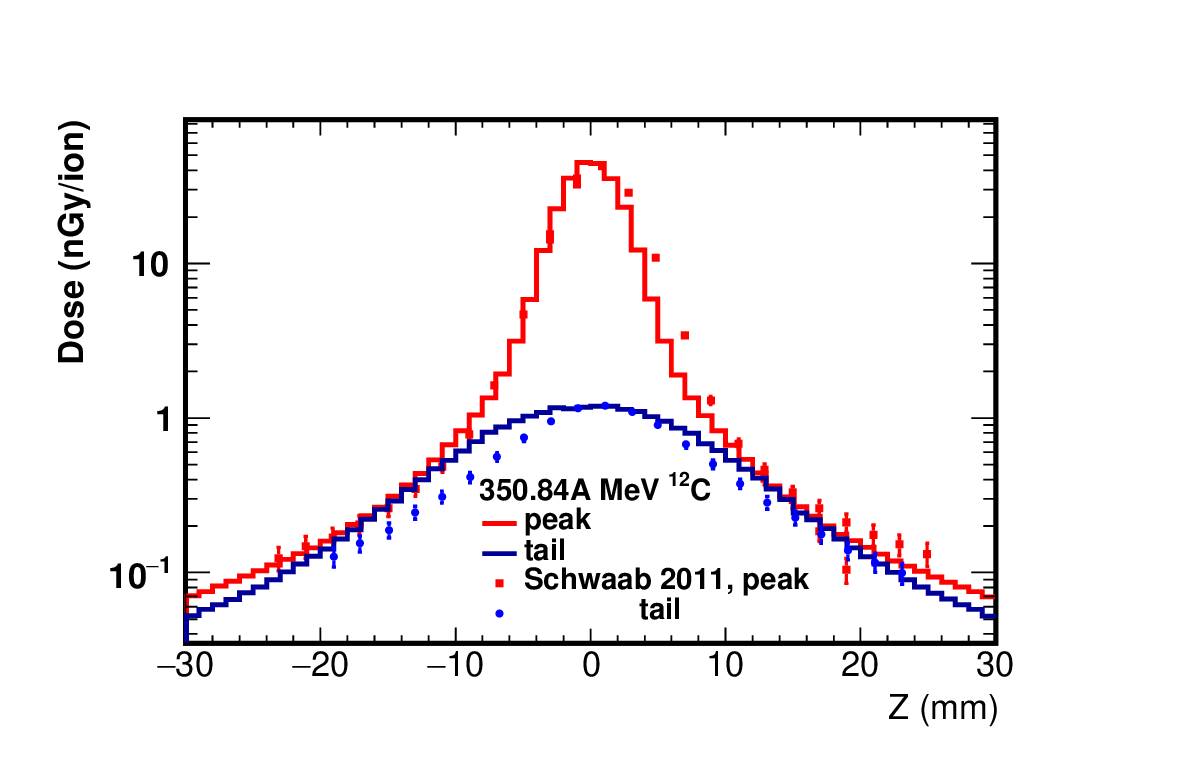}
    \end{minipage}
    \caption{Lateral profiles of  $182.66$~MeV proton beam (left) and $350.84$A~MeV $^{12}$C  beam (right) at the plateau (magenta), peak (red) and tail (blue) of the corresponding depth-dose distributions. The histograms of the respective colors represent Geant4 modeling. The measured lateral profiles~\cite{Schwaab2011} are shown by points.}
    \label{fig:lateral_profile}
\end{figure}

The lateral dose profiles of $182.66$~MeV proton and $350.84$A~MeV $^{12}$C beams calculated per beam particle are compared with data~\cite{Schwaab2011} in Fig.~\ref{fig:lateral_profile}. In this comparison, the maximum dose values at the centers of the lateral profiles were assumed to be the same in the measurements and calculations because a good agreement between the depth-dose profiles analyzed previously in Ref.~\cite{Pshenichnov2024} has been obtained in these two cases.  As can be seen from Fig.~\ref{fig:lateral_profile}, the measured lateral profiles of $^{12}$C beam are described well by calculations. A satisfactory agreement was also obtained for protons within 10~mm from beam axis. At the same time, the doses at the proton Bragg peak far from the beam axis, at $\sim 20$~mm, are underestimated  by a factor of 1.5--2. The disagreement between the measurements and calculations for protons at the plateau (up to a factor of 10 at $\sim 20$~mm from the beam axis) is apparently caused by neglecting the beam halo in our modeling. As stated~\cite{Schwaab2011}, the initial beam profile in air was affected by large angle scattering in the beam-monitoring systems, but their parameters were not reported. However, some underestimation of multiple scattering by Geant4 can not be excluded as well.  Indeed, the lateral profiles of pencil-like beams of protons calculated with a similar Geant4 version 10.2 were found to be narrower in comparison to measurements and other transport models, as reported in Ref.~\cite{Solie2017}. It can also be noted that the doses at very large distances (20--100~mm) from the axis of a 300A MeV $^{12}$C beam propagating in water are mainly delivered by various secondary particles rather than by multiple-scattered $^{12}$C~\cite{Burigo2013}. A good description of the dose values far from the beam axis has been demonstrated with Geant4 modeling in Ref.~\cite{Burigo2013}, and the same physics models were used in the present study. It can be concluded that detailed experimental studies of the lateral beam profiles, including the beam halo, are necessary for all considered projectiles in order to tune the models and obtain correct dose distributions in the valleys. At present, experimental data on the lateral profiles of ion minibeams are lacking.

\section{Approximating functions for dose profiles of mini- and pencil-like beams}\label{sec:approxmating_functions}

During the propagation of a proton or a nucleus in tissues, a sequence of events of individual elastic Coulomb scattering on atomic nuclei occurs, resulting in an angular deflection and a lateral displacement of the particle track~\cite{Regler2000}. The cumulative effect of multiple Coulomb scattering is expressed in a central core and a peripheral tail of the angular distribution of the particles after their propagation in tissues. The exact multiple scattering distributions are obtained by repeated convolution of the single scattering distribution, see Refs.~\cite{Regler2000,Fruhwirth2000} and references therein. The long tail of this distribution is associated either with very rare single scattering events at relatively large deflection angles or with elastic hadronic scattering on nuclei, and also with the production of secondary particles in nuclear reactions. Together with multiple Coulomb scattering, all these processes were taken into account in our Geant4 modeling.   

As shown~\cite{Fruhwirth2000}, the quantitative modeling of the core and tails of the angular distribution of a single particle deflected by multiple Coulomb scattering is possible by the sum of two Gaussian distributions with two essentially different standard deviations, $\sigma_2 \gg \sigma_1$.  Since the deflection angle $\theta$ resulting from multiple scattering is generally small, it is linearly translated to the radial displacement $r$ of the particle from its initial direction. It can be assumed that the dose in a certain voxel is proportional to the number of displaced particles passing through this voxel. Therefore, one can also try to approximate by a double-Gauss function the  radial dependence of the dose from a minibeam of infinitesimally small diameter calculated at a given depth by MC modeling.

For pencil-like beams of a few mm FWHM, the dose distribution from a single beam of protons or light nuclei of initial energy $E$ propagating along the $X$ axis in water can generally be characterized by a product of two Gaussians with the respective transverse dispersions, $\sigma_y(x)$ and $\sigma_z(x)$:
\begin{equation}
 D(E,x,y,z)=\frac{N_0(x)}{2\pi\sigma_y(x)\sigma_z(x)} \exp 
\left(-\frac{y^2}{2\sigma^2_y(x)}\right) \exp 
\left(-\frac{z^2}{2\sigma^2_z(x)}\right) .
\label{eq:plb-xyz}
\end{equation}
The energy $N(x)$ deposited per unit depth in a given water slab at the depth~$x$, as well as the dispersions $\sigma_y(x)$ and $\sigma_z(x)$ are specific for each kind of beam ion and depend on the initial beam energy $E$. In the case of the axial symmetry $\sigma_y(x)=\sigma_z(x)=\sigma(x)$ of the dose distribution from an individual pencil-like beam of protons or $^{12}$C used in active beam scanning~\cite{Bellinzona2015,Schwaab2011,Parodi2013}, the radial distance $r=\sqrt{y^2+z^2}$ from the central beam axis is introduced and Eq.~(\ref{eq:plb-xyz}) is rewritten as: 
\begin{equation}
 D(E,x,r)=\frac{N_0(x)}{2\pi\sigma^2(x)} \exp 
\left(-\frac{r^2}{2\sigma^2(x)}\right) .
\label{eq:plb-radial}
\end{equation}
In the following, the dependencies of $D$, $N_0$ and various parameters of the radial distributions on the depth $x$, the beam energy $E$ and the ion type are implied, but not explicitly written in order to simplify the notations.

It is worthwhile to note that the functions approximating dose profiles of pencil-like beams~\cite{Bellinzona2015,Schwaab2011,Parodi2013} typically contain the radial dependence of dose introduced in Eq.~(\ref{eq:plb-radial}). However, lateral dose profiles are usually measured~\cite{Schwaab2011} by placing detector arrays at a given distance $z$ from the beam axis and moving them perpendicular to it to obtain projected distributions of dose shown, in particular, in Fig.~\ref{fig:lateral_profile}. The general relations between radial and lateral dose distributions have been presented in Ref.~\cite{Embriaco2017} following Ref.~\cite{Papoulis1968}.

It has been shown~\cite{Schwaab2011,Parodi2013} that a double-Gaussian function characterized by the standard deviations, $\sigma_1$, $\sigma_2$ and by the weight $w$:
\begin{equation}
    D = N_0 \left\{\frac{(1 - w)}{2\pi \sigma_1^2} \exp 
\left(-\frac{r^2}{2\sigma_1^2}\right)+\frac{w}{2\pi \sigma_2^2} \exp \left( -\frac{r^2}{2\sigma_2^2} \right) \right\}
    \label{eq:double_gauss}
\end{equation}
provides certain improvements in describing  calculated and measured lateral dose profiles of pencil-like beams of protons and $^{12}$C with respect to a single Gaussian function. 
In Eq.~(\ref{eq:double_gauss}) the parameter $\sigma_1$ represents the standard deviation of the first narrower Gaussian associated with the profile of the initial beam, while the standard deviation $\sigma_2 > \sigma_1$ is a parameter of the second wider Gaussian associated with the dose delivered by particles scattered or produced at larger angles, with a larger displacement from the central beam axis. The parameter $w$ represents the weight assigned to the wider Gaussian. 

In Ref.~\cite{Bellinzona2015} the best description of the lateral dose profiles of pencil-like proton beams was obtained with the triple-Gaussian and double-Gaussian Lorentz-Cauchy functions, both of which have five parameters, but good results were also reported for the Gauss-Rutherford function, which has only three parameters.

In our study, Eq.~(\ref{eq:double_gauss}) was supplemented by an additional term to account for particles scattered in rare single scattering events at very large angles and products of nuclear reactions. Its form is motivated by the dependence of the differential cross section of Rutherford scattering on nuclei taking into account the screening effect of electrons at large impact parameters, as described in Ref.~\cite{Fruhwirth2000}. This third term will be denoted below as the Rutherford contribution to the radial dose distribution. It contains a single parameter $b$ to describe the screening effect and remove the singularity at very low scattering angles typical for the original Rutherford scattering formula~\cite{Fruhwirth2000}. Therefore, the sum of the three terms was used in our study as an alternative to the double-Gauss parametrization of the doze profile: 
\begin{equation}
    D = N_0 \left\{ \frac{(1 - w_1 - w_2)}{2\pi \sigma_1^2} \exp 
\left(-\frac{r^2}{2\sigma_1^2}\right)+\frac{w_1}{2\pi \sigma_2^2} \exp \left( -\frac{r^2}{2\sigma_2^2} \right) +  \\
      \frac{w_2 \cdot 2 b r}{\pi^2(r^2 + b^2)^2} \right\} \ .
\label{eq:double_gauss_rutherford}
\end{equation}

It can be noted that in Eq.~(\ref{eq:double_gauss_rutherford}) a slightly different polynomial form of the Rutherford contribution to the radial dose profile is employed with respect to a similar contribution to the lateral dose distribution considered in Ref.~\cite{Bellinzona2015}.    
In the following the parametrizations given by Eqs.~(\ref{eq:double_gauss}) and (\ref{eq:double_gauss_rutherford}) are denoted as double-Gaussian (DG) and double-Gaussian-Rutherford (DGR) parametrizations, respectively. A common fitting procedure is employed with the both DGR and DG parametrizations, but the weight of the third Rutherford term $w_2$ is set to zero in the case of the DG parametrization.

\section{Fit of dose profiles of single minibeams}\label{sec:single_minibeams}

Dose profiles of single minibeams of protons, $^4$He, $^{12}$C and  $^{16}$O nuclei with initial FWHM of $0.3$~mm and $0.5$~mm propagating in water were calculated  with Geant4 using the same physics lists as for pencil-like beams, see Sec.~\ref{sec:validation}. The initial radial minibeam profiles at the entrance to the phantom were assumed to be Gaussian.  As pointed out in Refs.~\cite{Zlobinskaya2013,Sammer2017,Sammer2021a}, the initial angular divergence of minibeams can be neglected in simulations. In Ref.~\cite{Mayerhofer2021} magnetically focused minibeams of protons were considered with non-negligible beam divergence. However, zero divergence at the entrance to tissues was also adopted as a possible option.

The minibeam kinetic energies ($152.7$~MeV for protons, $152.7A$~MeV for $^4$He, $290.0A$~MeV for $^{12}$C and $345.4A$~MeV for $^{16}$O, with negligible energy spread) were chosen to provide the positions of the Bragg at the depth of 160~mm  to facilitate the comparison of the doses delivered by all four projectiles. In order to make the notations shorter, these energies are denoted below as $152$~MeV, $152A$~MeV, $290A$~MeV and $345A$~MeV, respectively. In Geant4 modeling minibeams were directed to the center of a rectangular water phantom with dimensions of $10\times 10 \times 200$~mm$^3$ along its longer central axis. For the purpose of the dose scoring the phantom was divided into concentric ring voxels with the bin size in depth $x$ of 0.025~mm and with the same bin size in radius $r$. The doses were calculated in nGy per beam particle at the end of each Monte Carlo run. Up to $2\times 10^6$ events were generated for each kind of projectiles and each beam diameter, and the 2D histograms representing dose variation with depth and distance from the beam axis were finally stored in $\ast$.root files after the end of each run for further analysis and visualization with CERN ROOT Data Analysis Framework~\cite{Brun1997,BrunSoft} of version 6.28/04.   

The $\chi^2$ least-squares minimization fit algorithm Migrad from Minuit library~\cite{James1975} implemented in CERN ROOT was used to fit a set of 1D histograms representing the radial dose distributions at a given depth. These 1D histograms were obtained as slices of the 2D histograms representing the dose distributions in the water phantom calculated with Geant4 modeling. In order to reduce the correlation between the fit parameters and, respectively, their uncertainties, the function given by Eq.~(\ref{eq:double_gauss_rutherford}) was temporarily substituted in the fitting procedure by the following function:
\begin{equation}
    D^\star = \frac{N_1}{2\pi} \exp \left(-\frac{r^2}{2\sigma_1^2} \right) + \frac{N_2}{2\pi} \exp \left(-\frac{r^2}{2\sigma_2^2} \right) + N_3 \frac{2 b r}{\pi^2(r^2 + b^2)^2} 
    \label{eq:dgr_fit} \ .
\end{equation}
After the completion of the fit for each slice the values of $\sigma_1$, $\sigma_2$, $b$, $N_1$, $N_2$, $N_3$ in Eq.~(\ref{eq:dgr_fit}) were extracted, and then they were converted to the parameters used in Eq.~(\ref{eq:double_gauss_rutherford}) via the following relations:  
$N_0 = \sigma_1^{2} N_1+\sigma_2^{2} N_2 + N_3$, $w_1 = \sigma_2^2 N_2/N_0$,  $w_2 = N_3/N_0$.

The parameters of the function $D$ characterizing the dose from a single minibeam in water were obtained step by step at many points in depth starting from the first entrance slice with its center placed at 0.9--2.6~mm depth depending on the type of beam particles. Similarly, different step sizes from 0.3~mm to 6~mm were implemented in the domains of plateau, peak and tail of the depth-dose distributions of the minibeams depending on their initial FWHM and the type of beam particles to ensure the stability of the fit.

\begin{figure}[htb!]
    \begin{minipage}{0.49\linewidth}
    \includegraphics[width = 1.05\linewidth]{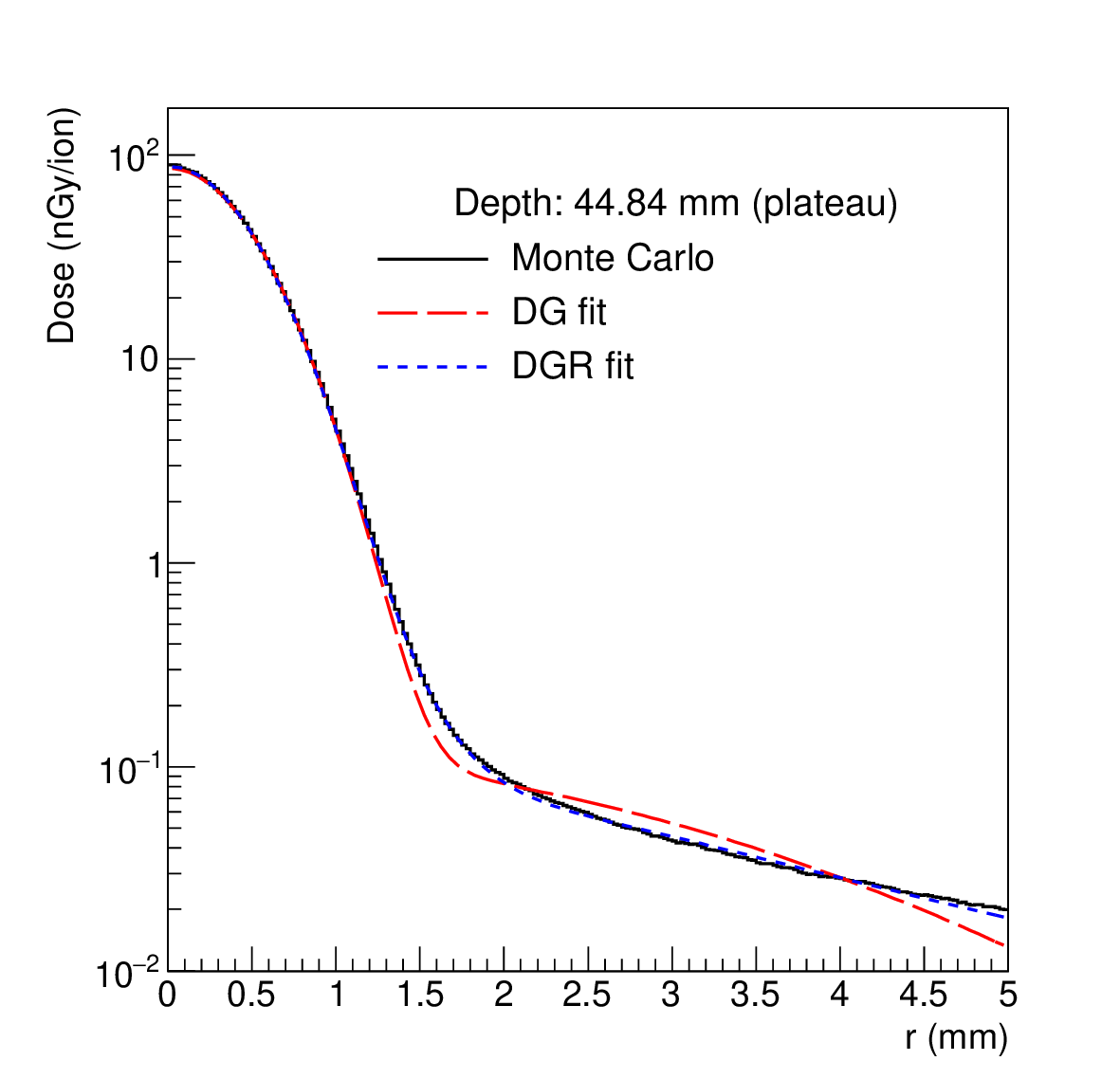}
    \end{minipage}
    \begin{minipage}{0.49\linewidth}
    \includegraphics[width = 1.05\linewidth]{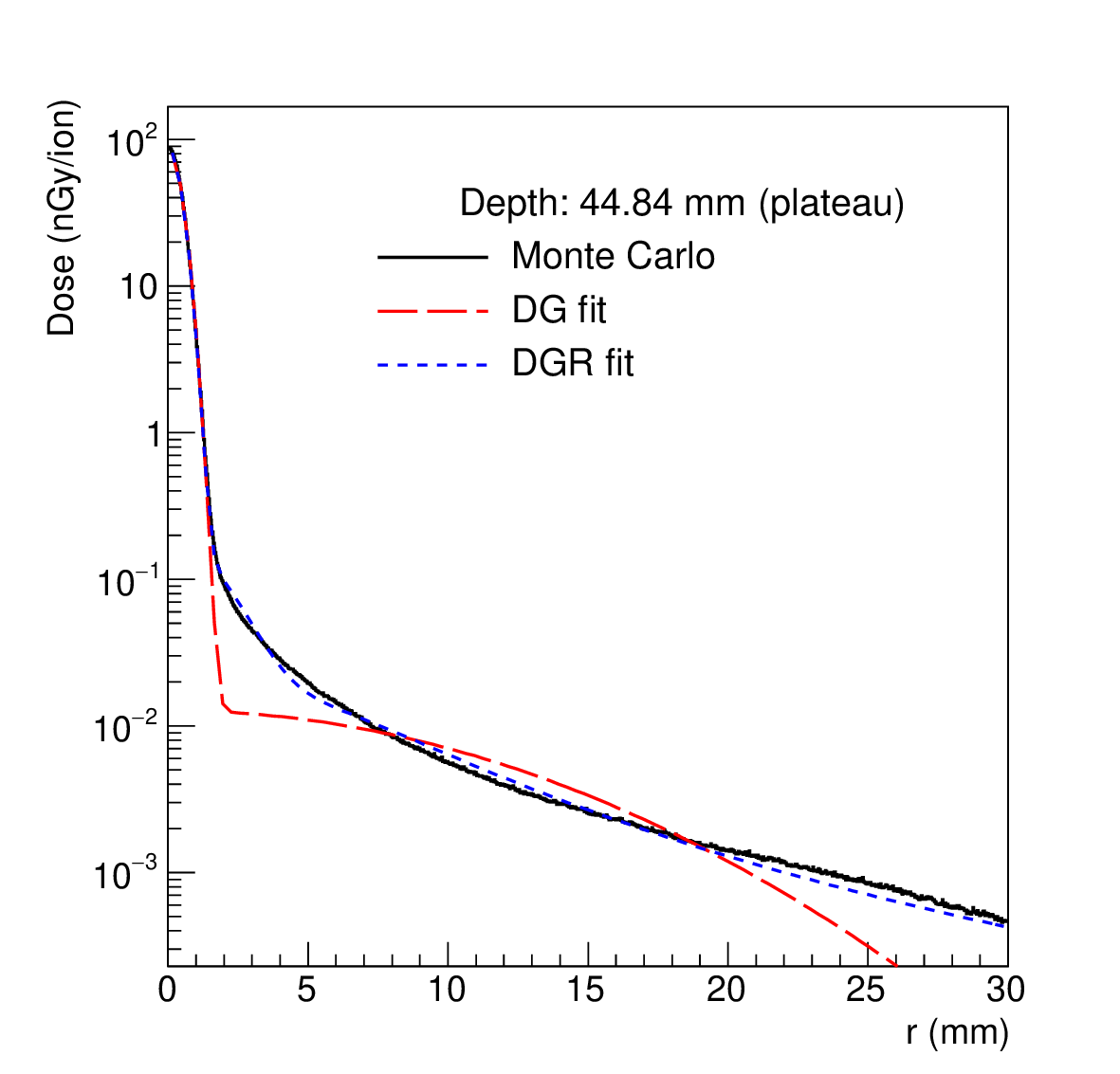}
    \end{minipage}     
    \caption{Radial distribution of dose from a 152~MeV proton minibeam of 0.5~mm FWHM obtained by Monte Carlo modeling (solid line histogram) at the depth of 44.84~mm in water. The approximations obtained with the DG and DGR functions are shown by dashed and dotted lines, respectively. The fits with both functions were performed up to radii of 5~mm (left) and 30~mm (right).}
    \label{fig:H1_152A_DG_vs_DGR}
\end{figure}

Examples of results of the fit with the DG and DGR functions for a radial profile of 152~MeV proton minibeam of 0.5~mm FWHM at the depth of 44.84~mm in water are shown in Fig.~\ref{fig:H1_152A_DG_vs_DGR}. As can be seen, at the plateau the DGR function provides a better description of the radial dose profile  up to 30~mm compared to the DG function previously used to approximate the dose profiles of pencil-like beams. The shape of a long tail of the radial dose profile of the minibeam in MC modeling is substantially different from the shape of the DG function at larger radii. Since similar difficulties in describing the radial profiles in the plateau regions with the DG function were  also found for minibeams of $^4$He, $^{12}$C and $^{16}$O, only the DGR approximations are considered in the following. 

At the beginning of each fit, the initial values of $\sigma_1$ and $N_1$, which describe the width of the initial beam core and the entrance dose, respectively, were chosen for the first slice taking into account the transverse minibeam dimension and the maximum dose at the entrance of the phantom. Much smaller values of $N_2$ and $N_3$ were assumed for the first slice, making the choice of $\sigma_2$ and $b$ less crucial. The requirement to be positive was applied to all fit parameters. The parameters of the fit function given by Eq.~(\ref{eq:dgr_fit}) obtained at each step were used as initial parameter values for the fit performed at the next step. This provided a sufficiently slow variation of the parameter set with depth, including the Bragg peak region, where the step size was reduced to 0.3~mm from its values of 3--6~mm at the plateau to maintain the smoothness of such variations. However, in the first iteration of the fit of the radial dose profile, based on $\sigma_1$ and $N_1$ of the initial beam, the quality of the fit was poor in the very first slice. Therefore, in the second iteration, the parameter values obtained for the second slice were used as initial parameters for the first slice. This finally provided a good fit quality from the entrance to the phantom till the Bragg peak.

\begin{figure}[tb!]
    \centering
    \begin{minipage}[c]{.49\textwidth}
        \includegraphics[width = 1.1\textwidth]{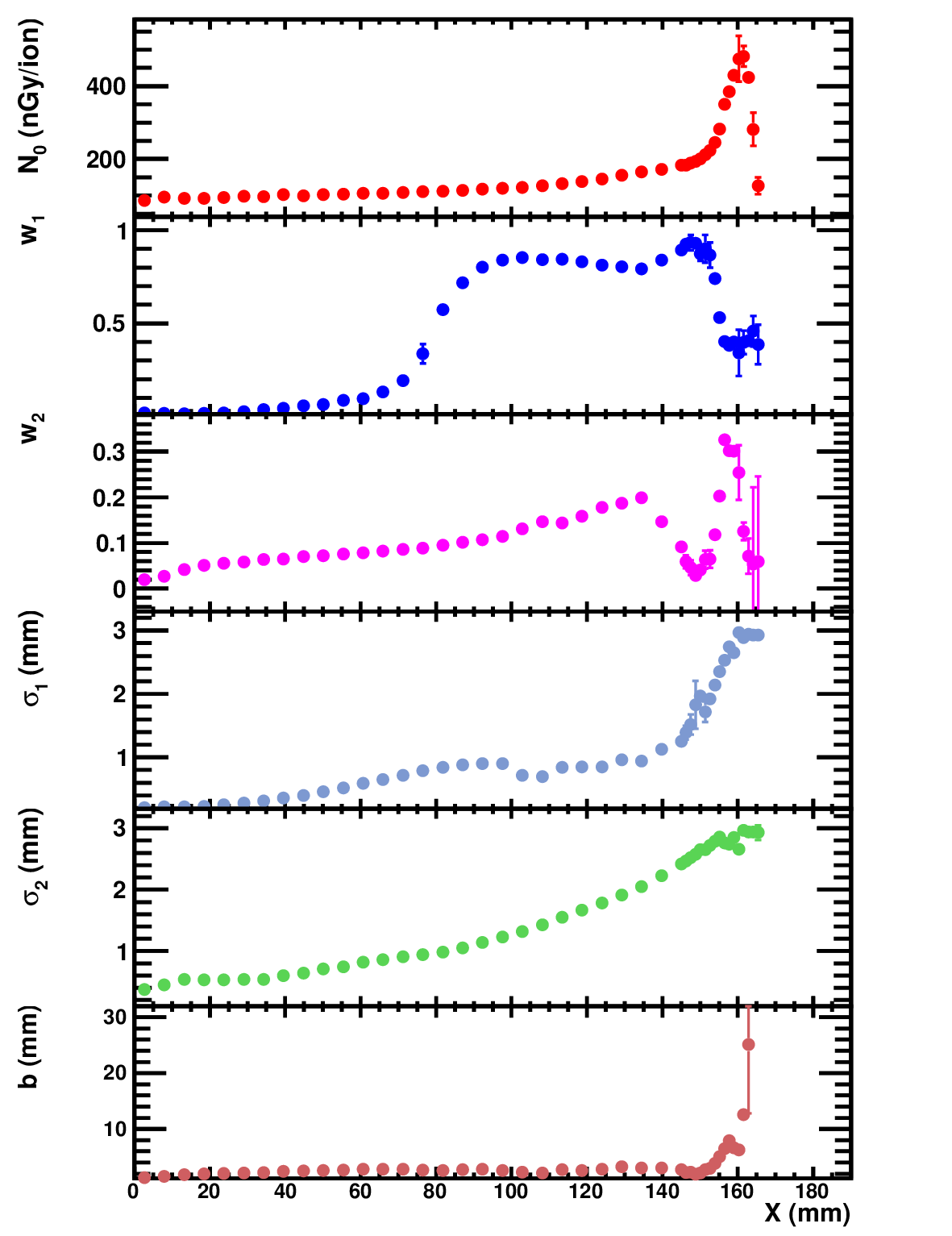} 
    \end{minipage}
    \begin{minipage}[c]{.49\textwidth}
        \includegraphics[width = 1.1\textwidth]{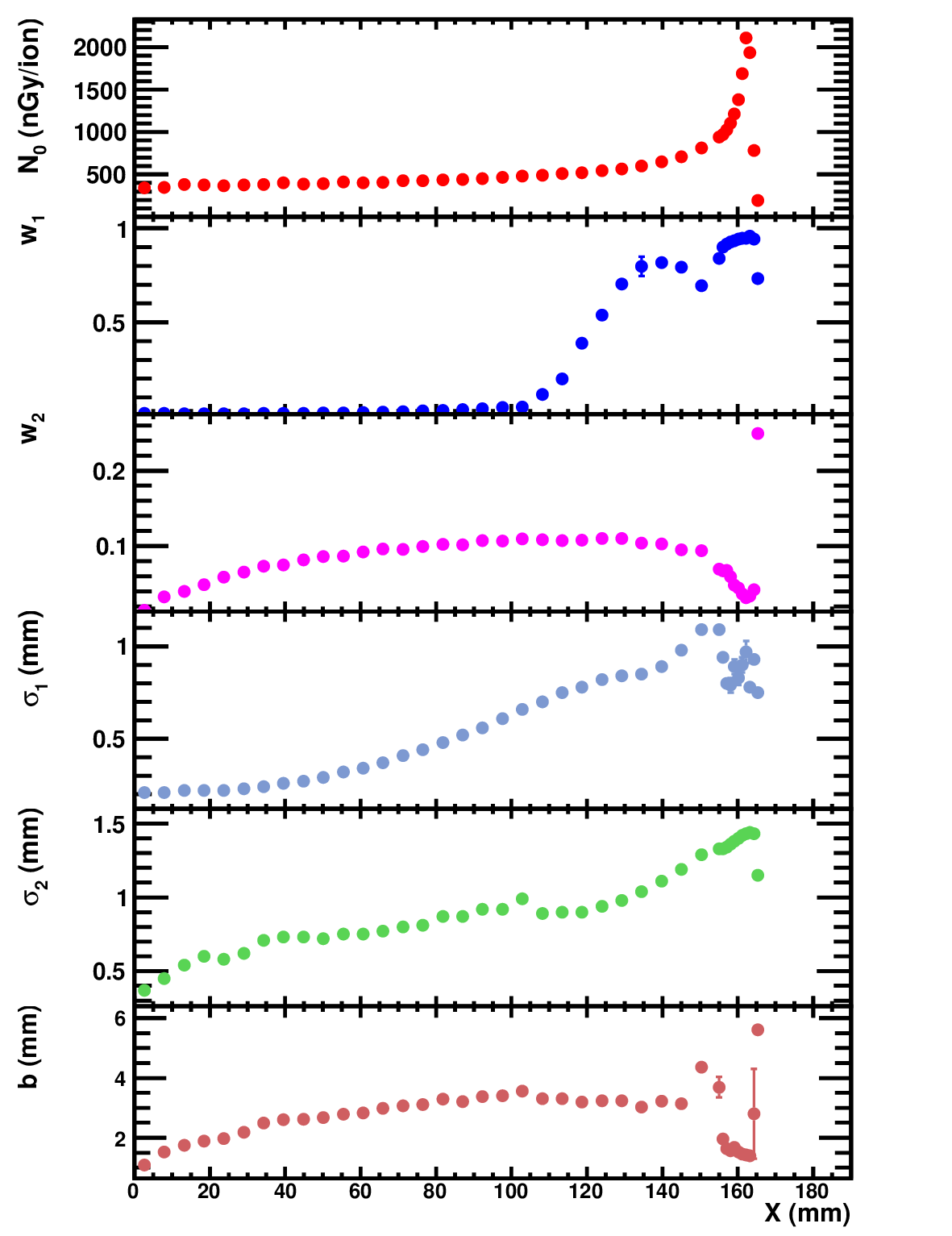}
    \end{minipage}
    \caption{Dependence of the parameters of the DGR approximating function on depth in water for proton (left) and $^4$He (right) minibeams of 0.5~mm FWHM.}
    \label{fig:hyd_hel_pars}
\end{figure}

The dependence on depth $x$ in water of the parameters $N_0$, $w_1$, $w_2$, $\sigma_1$, $\sigma_2$ and $b$ obtained by fitting the dose distributions of $152$~MeV proton and $152A$~MeV $^4$He minibeams of 0.5~mm FWHM with the DGR function, Eq.~(\ref{eq:double_gauss_rutherford}), is shown in Fig.~\ref{fig:hyd_hel_pars}. The fit procedures with the DGR function successfully converged at all points in depth before the Bragg peak and a few millimeters beyond it. However, very low dose values and their large statistical errors calculated in the tail region prevented accurate determination of the DGR parameters far beyond the peak. As can be seen from Fig.~\ref{fig:hyd_hel_pars}, for the lightest projectiles considered in this work, $\sigma_1$, $\sigma_2$ and $b$ increase slowly with depth due to the multiple scattering of primary beam particles and the production of secondary particles (mostly protons and neutrons) in nuclear reactions in water. The contribution of the Rutherford term also increases with depth, but remains small compared to the contribution of the first Gaussian. 

\begin{figure}[tb!]
    \centering
    \includegraphics[width = 1.05 \textwidth]{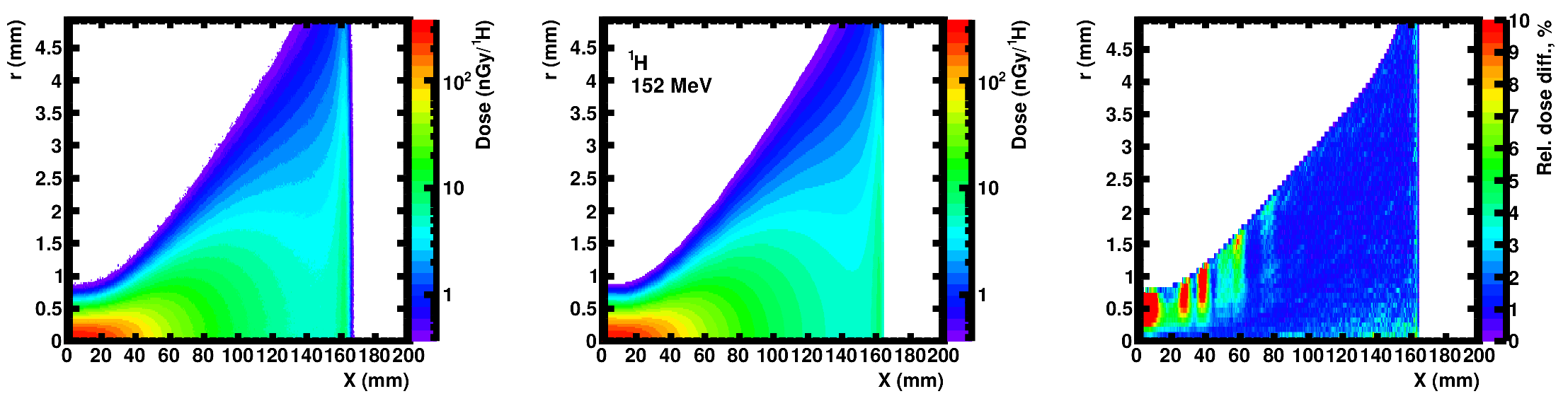}
    \includegraphics[width = 1.05\textwidth]{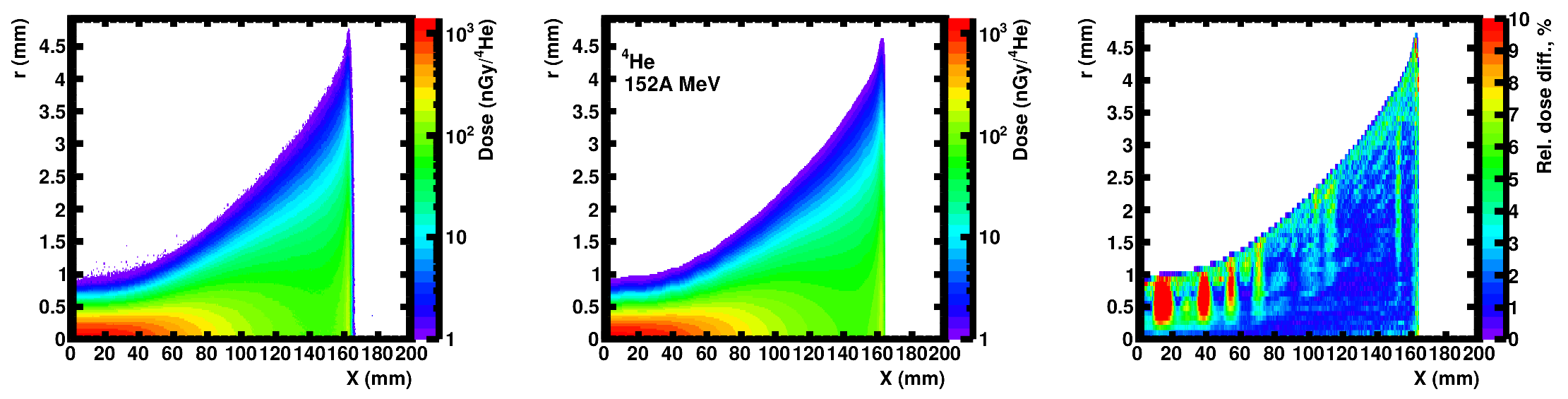}
    \caption{Distributions of dose (nGy/ion) in water for $152$~MeV proton (top row) and $152A$~MeV $^4$He (bottom row) minibeams of $0.5$~mm FWHM. The distributions obtained directly from Geant4 modeling, calculated with the DGR function with its parameters resulted from the fit, and the modulus of their relative difference (in \%)  are shown, respectively, from left to right as functions of depth and distance from the beam axis.}
    \label{fig:hyd_hel}
\end{figure}

\begin{figure}[tb!]
    \centering
    \includegraphics[width = 1.05 \textwidth]{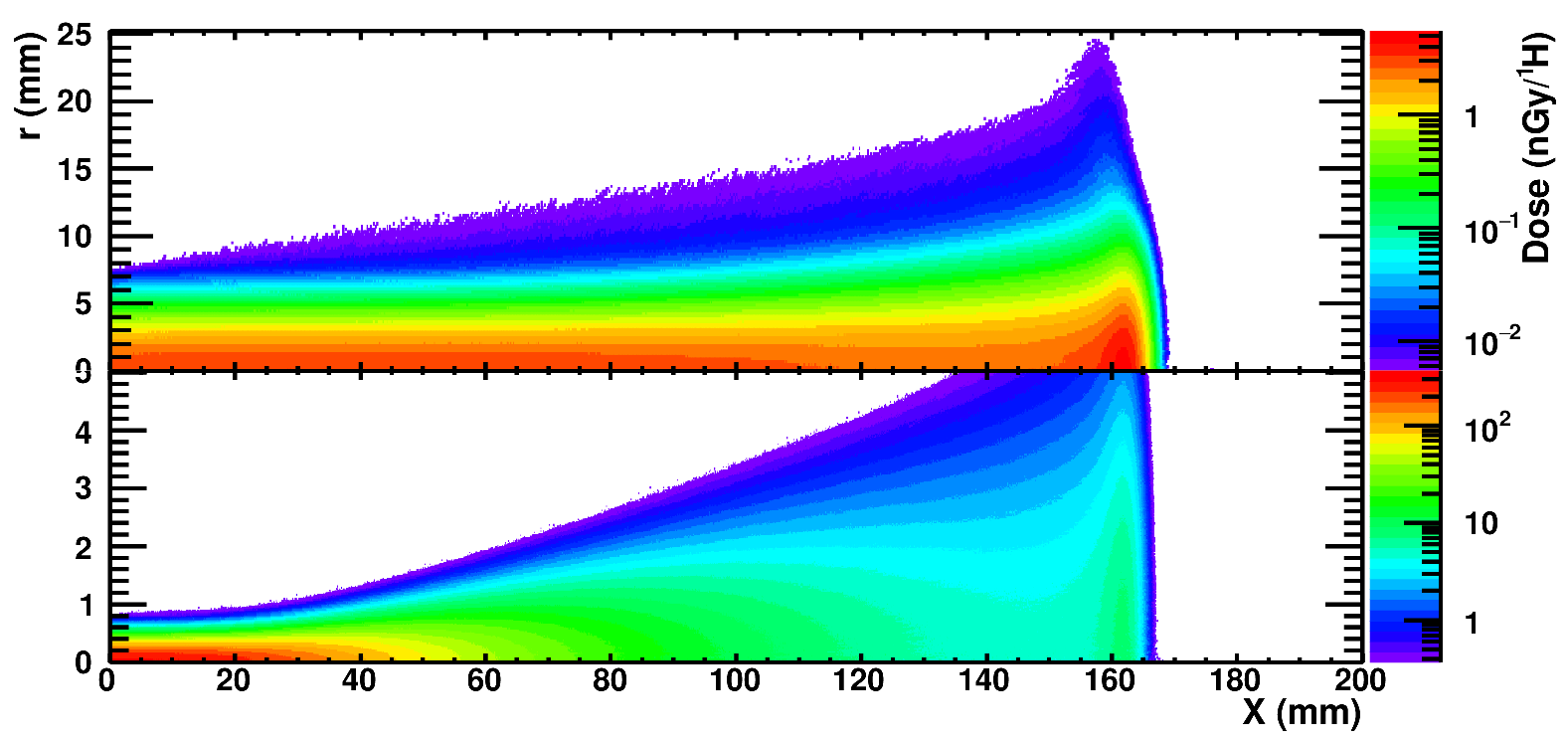}
    \caption{Distributions of dose (nGy/ion) in water for $152$~MeV proton beams of 5~mm (top) and 0.5~mm (bottom) FWHM. }
    \label{fig:pb_vs_mini}
\end{figure}

The evolution of the fit parameters seen in Fig.~\ref{fig:hyd_hel_pars} is due to a specific shape of the dose distribution of minibeams in water shown in Fig.~\ref{fig:hyd_hel} as a function of depth and radius. This shape is drastically  different from that calculated for a 152~MeV pencil-like beam of 5~mm FWHM in water, see Fig.~\ref{fig:pb_vs_mini}. There is a significant difference between the distributions of the dose on the beam axis in these two cases. Many protons of the pencil-like beam are deflected away from the beam axis due to multiple scattering, but the dose on the beam axis at the Bragg peak still remains comparable to the axis dose at the phantom entrance. In contrast, the dose on the axis of the proton minibeam is reduced by a factor of 10--20 already in the first 40\% of the range of protons in water. This is because almost all minibeam protons are deflected laterally well beyond the initial  FWHM. As a result, the dose on the beam axis at the Bragg peak is only $\sim 1$\% of the entry axis dose. 

As can be seen from Fig.~\ref{fig:hyd_hel}, the dose distributions, $D_{\mathrm{MC}}$, calculated with Geant4 and with the DGR functions, $D_{\mathrm{Fit}}$, with the parameters obtained from the fit presented in    Fig.~\ref{fig:hyd_hel_pars}, have identical shapes. In order to quantitatively access the fit quality, the modulus of the relative difference 
\begin{equation}
    \delta = 2\times \frac{D_{\mathrm{MC}}-D_{\mathrm{Fit}}}{D_{\mathrm{MC}}+D_{\mathrm{Fit}}}
    \label{eq:delta}
\end{equation}
between these two dose distributions was calculated at different locations in the phantom, see Fig.~\ref{fig:hyd_hel}. As found, at most locations $|\delta|\leq 3$\%, while $|\delta|$ reaches 10\% only in a very limited number of voxels. As can be seen in Fig.~\ref{fig:hyd_hel}, $|\delta|$ increases up to 10\% at low phantom depths at the periphery of the proton and helium beams. There, the fitting procedure becomes less accurate in describing the dose at the tail of the radial dose distribution because it is 10$^2$ - 10$^3$ times lower than the dose at the beam axis. In particular, some differences between MC and the fit can be traced in Fig.~\ref{fig:H1_152A_DG_vs_DGR} only at the tail of the radial distribution, but not near the beam axis

\begin{figure}[tb!]
    \centering
    \begin{minipage}[c]{.49\textwidth}
        \includegraphics[width = 1.1\textwidth]{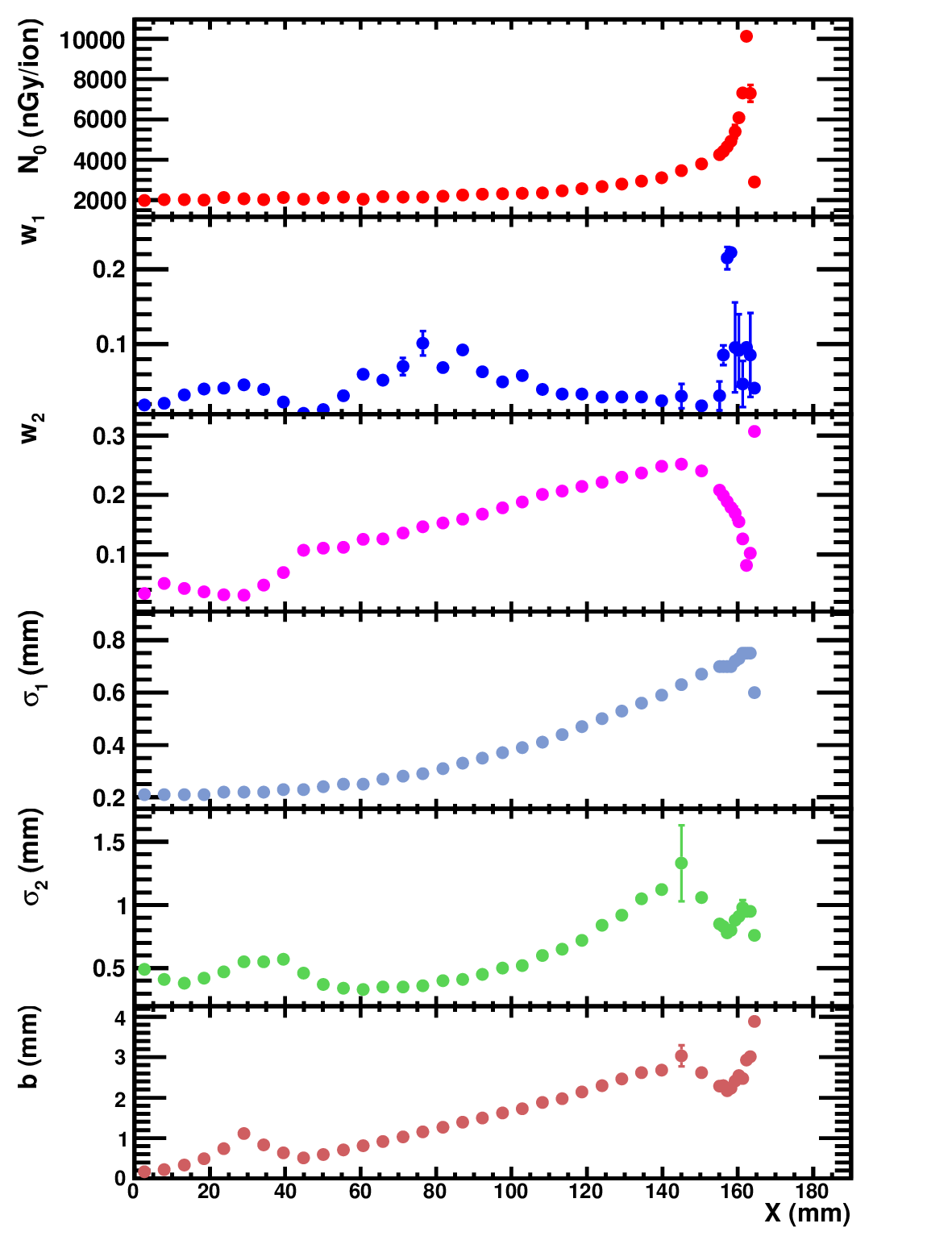} 
    \end{minipage}
    \begin{minipage}[c]{.49\textwidth}
        \includegraphics[width = 1.1\textwidth]{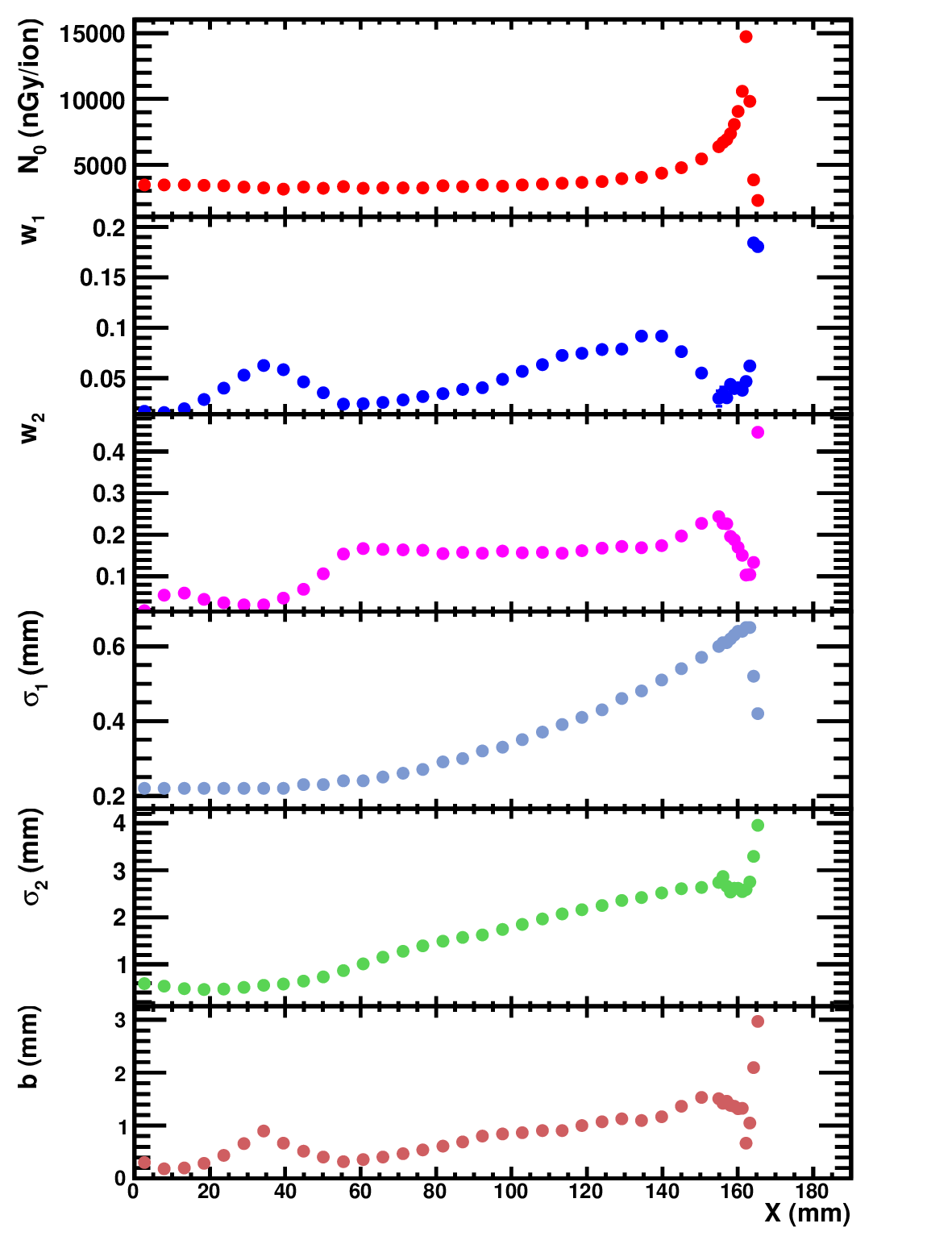}
    \end{minipage}
    \caption{Same as in Fig.~\protect\ref{fig:hyd_hel_pars}, but for $290A$~MeV $^{12}$C (left) and $345A$~MeV $^{16}$O (right) minibeams of 0.5~mm FWHM.}
    \label{fig:carb_oxy_pars}
\end{figure}

The dependence on depth $x$ in water of the parameters $N_0$, $w_1$, $w_2$, $\sigma_1$, $\sigma_2$ and $b$ obtained from the fit of the dose distributions of $290A$~MeV $^{12}$C and $345A$~MeV $^{16}$O minibeams of 0.5~mm FWHM with the DGR function, Eq.~(\ref{eq:double_gauss_rutherford}), is shown in Fig.~\ref{fig:carb_oxy_pars}. The fit procedures with the DGR function successfully converged at all points in depth before the Bragg peak and a few millimeters beyond it. However, low dose values and their large statistical errors calculated in the tail make less accurate the determination of the DGR parameters beyond the peak. As can be seen from Fig.~\ref{fig:carb_oxy_pars}, for minibeams of $^{12}$C and $^{16}$O, $\sigma_1$, $\sigma_2$ and $b$ also increase slowly with depth due to the multiple scattering of primary beam particles and the production of secondary particles (protons, neutrons and light nuclear fragments) in nuclear reactions in water. Nevertheless, in contrast to proton and $^4$He minibeams, the contribution of the Rutherford term, $w_2$ is twice as large important because of more frequent production of secondary particles in nuclear reactions induced by  $^{12}$C and $^{16}$O. 

\begin{figure}[tb!]
    \centering
    \includegraphics[width = 1.05\textwidth]{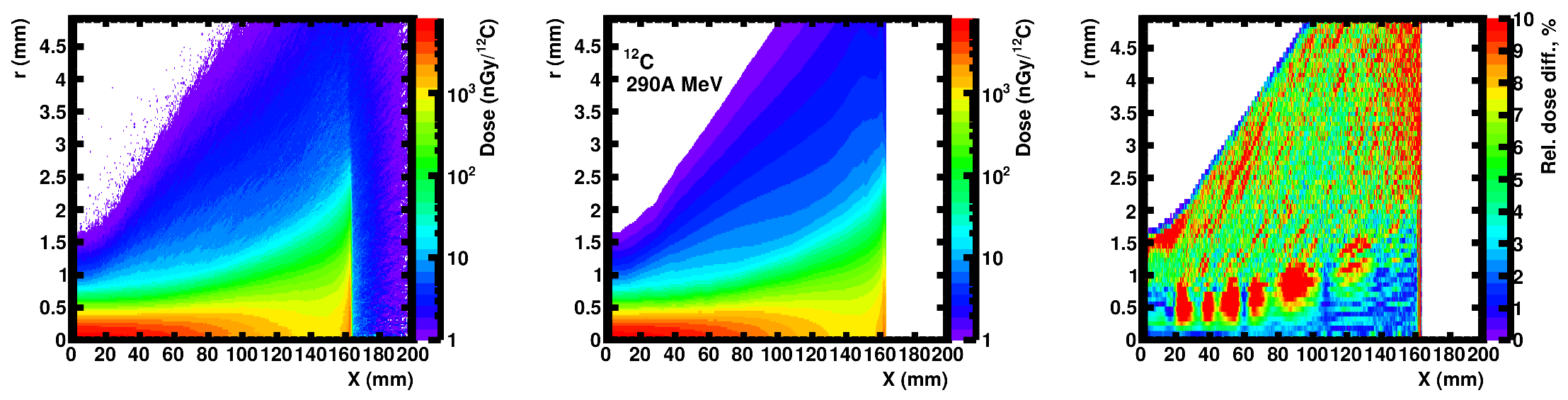}
    \includegraphics[width = 1.05\textwidth]{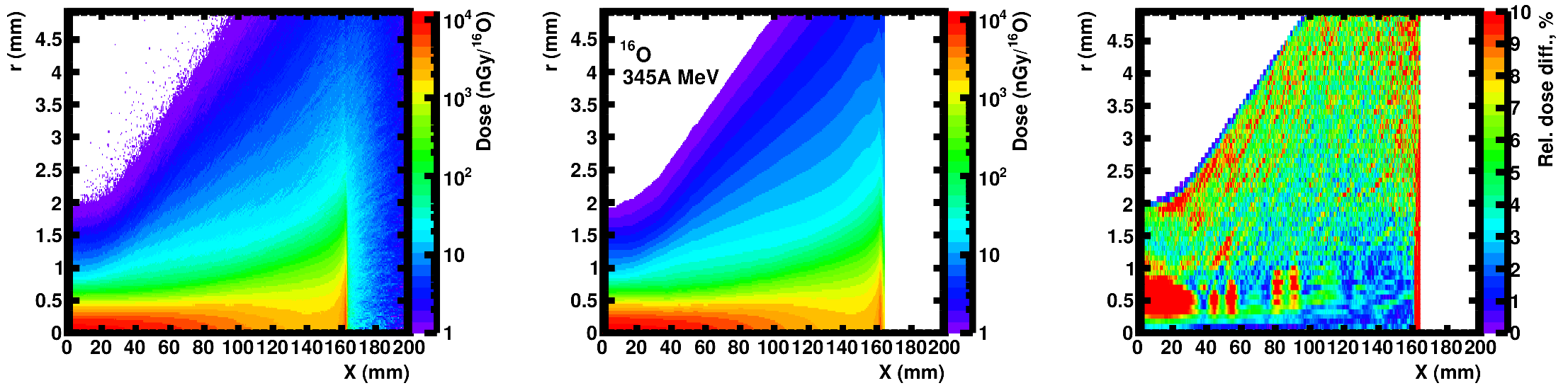}
    \caption{Same as in Fig.~\protect\ref{fig:hyd_hel}, but for $290A$~MeV $^{12}$C (top row) and $345A$~MeV $^{16}$O (bottom row) minibeams of 0.5~mm FWHM.}
    \label{fig:oxy_carb}
\end{figure}

The distributions of dose from minibeams of $^{12}$C and $^{16}$O were also calculated using Geant4 and the DGR functions with their parameters obtained from the fit, see Fig.~\ref{fig:carb_oxy_pars}. These dose distributions are shown in Fig.~\ref{fig:oxy_carb}, and they demonstrate identical shapes. In order to assess their similarity quantitatively, the moduli of $\delta$ given by Eq.~(\ref{eq:delta}) were calculated and presented in Fig.~\ref{fig:oxy_carb}. The values of $|\delta|$ are mostly under 6\%, while at some points $|\delta|\sim 10$\%. Similarly to the case of proton and $^4$He minibeams, the fit quality for $^{12}$C and $^{16}$O is reduced at low phantom depth and far from the beam axis. Due to the relatively modest lateral deflection of heavy projectiles, $^{12}$C and $^{16}$O, in comparison to protons and $^4$He, the reduction of dose on the beam axis close to the Bragg peak is less prominent, as shown in Fig.~\ref{fig:oxy_carb}. 

However, as shown by calculations, the general shape of the dose distributions of $^{12}$C and $^{16}$O minibeams remains very different from that of $^{12}$C and $^{16}$O pencil-like beams of the same energy. This difference further justifies the suggestion to use the DGR parametrization for minibeams, a function that has never been considered~\cite{Bellinzona2015,Schwaab2011,Parodi2013} for pencil-like beams.

In summary, for single mini\-beams of protons, $^4$He, $^{12}$C and $^{16}$O with 160~mm range in water, the modulus of the relative difference $\delta$ between the dose distributions calculated with Geant4 and based on approximations with the DGR functions is mostly under 6\%. Only in a very limited number of voxels $|\delta|$ reaches 10\%. Taking into account the dose variations at the level of three orders of magnitude between the beam axis and its far periphery ($r\sim 5$~mm), the DGR functions can be accepted as relevant approximations of the dose profiles of single minibeams. Below, in Secs.~\ref{sec:PVDR16} and \ref{sec:DVH16}, the quality of the fit is also evaluated in terms of its ability to reproduce the PVDR and DVH calculated for arrays of minibeams.

\section{Dose profiles of minibeam arrays}\label{sec:minibeam_arrangements}

The generation of minibeams and their implementation in a clinical environment is a complicated task~\cite{Schneider2022}. Arrays of minibeams can be generated either by propagating conventional pencil-like beams through multi-hole mechanical collimators~\cite{Charyyev2020} or directly by advanced magnetic focusing~\cite{Schneider2021a}. The focusing of minibeams in a scanning approach is considered superior to the collimation technique due to reduced beam halos and higher PVDRs~\cite{Datzmann2020}. The conceptual design of dedicated facilities based on linear proton accelerators have recently been proposed~\cite{Mayerhofer2021,Schneider2021a} for preclinical studies. However, it is still premature to assume that the parameters of magnetically focused minibeam arrays are well established even for protons. Therefore, in the present work, arrays of parallel circular minibeams with the same geometry for protons, $^4$He, $^{12}$C and $^{16}$O and negligible initial divergence were considered to facilitate the comparison of simulation results obtained for these types of projectiles.

\begin{figure}[tb!]
    \centering
    \begin{minipage}[c]{.49\textwidth}
        \includegraphics[width = 1.\textwidth]{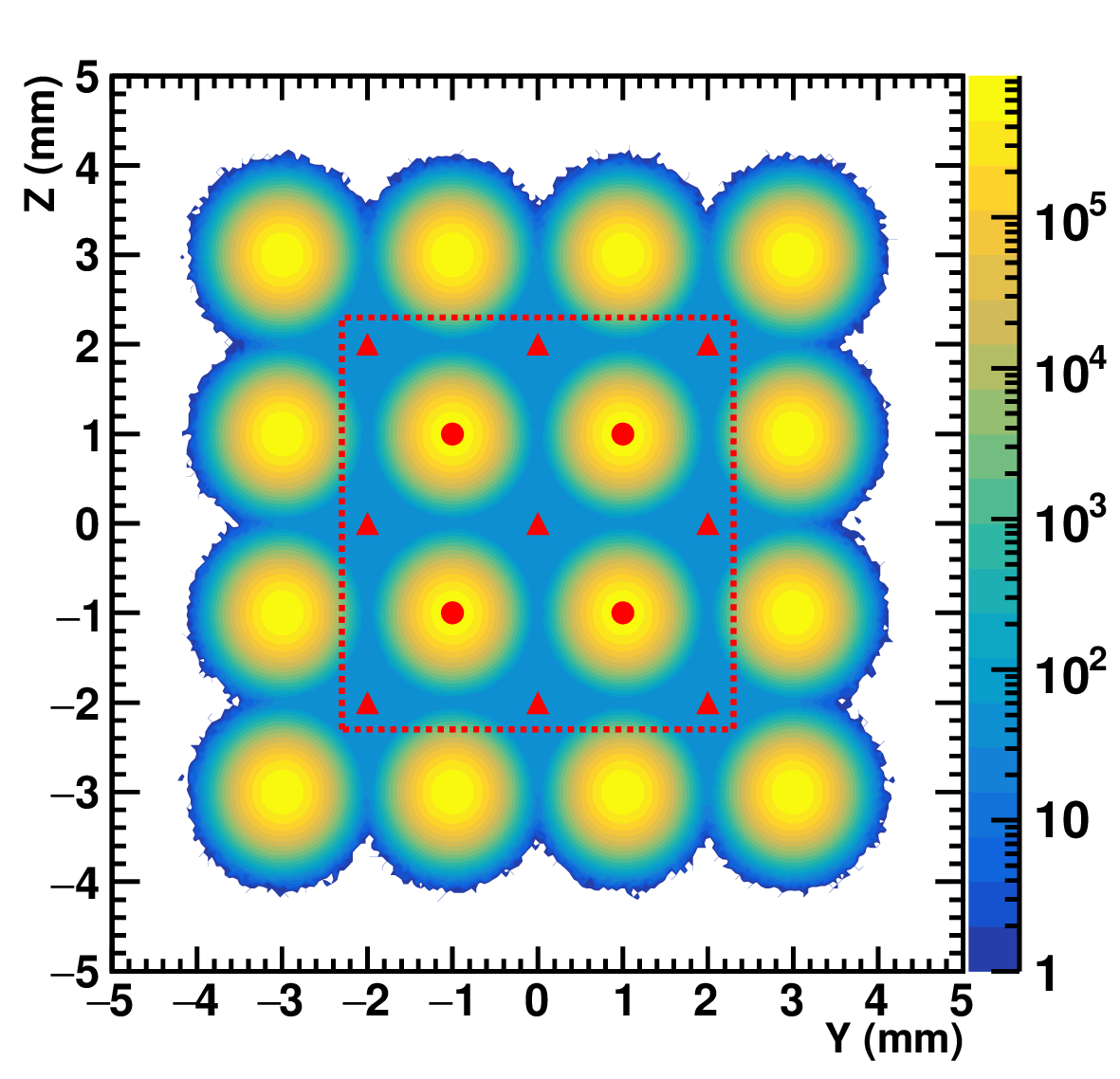} 
    \end{minipage}
    \begin{minipage}[c]{.49\textwidth}
        \includegraphics[width = 1.\textwidth]{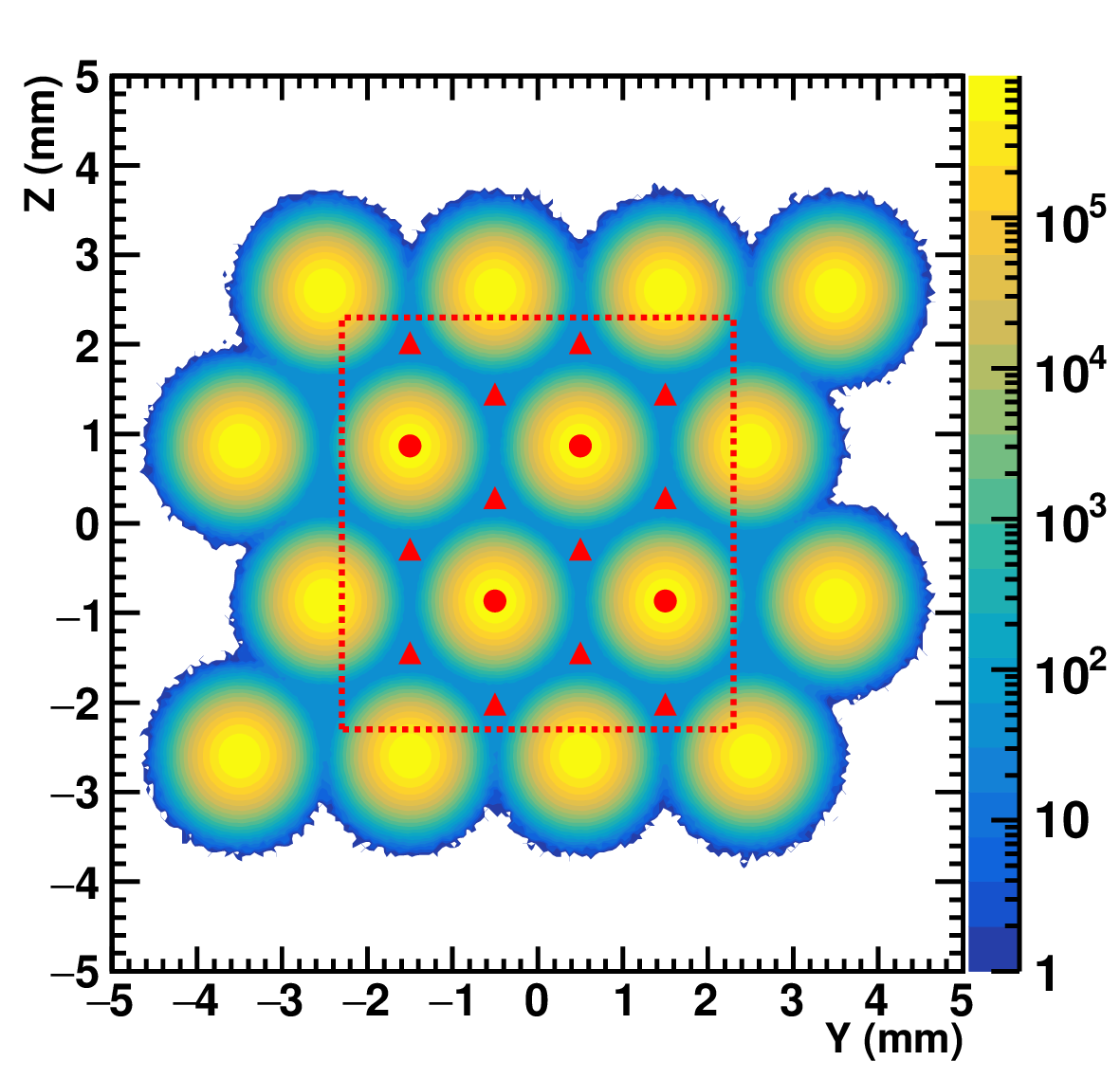}
    \end{minipage}
    \caption{Transverse distributions of beam intensity (arbitrary units) at the entrance to the water phantom for arrays of 16 minibeams of 0.5~mm FWHM. The beam centers are placed according to either a  rectangular (left) or hexagonal (right) grid. The positions of the peak and valley dose values used in the PVDR calculations are marked by circles and triangles, respectively. The boundaries of the scoring volumes used in the DVH calculations are indicated by dashed lines.}
    \label{fig:beam_intensity_transverse}
\end{figure}

The phantom of the same size ($10\times 10\times 200$~mm$^3$) as in the MC modeling with a single minibeam described in Sec.~\ref{sec:single_minibeams}, was used in calculations of the distributions of dose from arrays of  $152$~MeV proton, $152A$~MeV $^4$He, $290A$~MeV $^{12}$C and $345A$~MeV $^{16}$O minibeams. In this modeling, the centers of 16 parallel mininibeams of 0.3~mm or 0.5~mm FWHM were placed at the entrance of the phantom according to either a rectangular or hexagonal grid to represent an elementary cell of a minibeam array, as shown in Fig.~\ref{fig:beam_intensity_transverse}.  The center-to-center distance between the beams was set to 2~mm for the both minibeam patterns, which are considered~\cite{Meyer2019,Sammer2017,Charyyev2020} for applications in minibeam therapy. As shown in Ref.~\cite{Sammer2017}, circular beams, in particular in the hexagonal arrangement, were superior to the planar grid geometry in minimizing normal tissue damage and maximizing the efficacy of spatial dose fractionation. Therefore, the present study focuses on the consideration of rectangular and hexagonal grid patterns of circular minibeams. 

Each minibeam array with its cross section of about $8\times 8$~mm$^2$  can replace a typical single pencil like beam of a few mm transverse size used in the conventional particle therapy.
The central axis of each minibeam was aligned parallel to the longest axis of the water phantom, which was divided into voxels of $0.1\times 0.1\times 0.1$~mm$^3$ size. The 3D dose distributions in Cartesian coordinates for the minibeam arrays were obtained at the end of the Monte Carlo modeling as 3D-histograms in ROOT files for each type of beam particle, minibeam FWHM (0.3/0.5~mm) and grid pattern (rectangular/hexagonal). Typically, up to $10^7$ histories of beam protons were simulated in each case, requiring up to a day of CPU time in a multithreading mode on 24 CPU cores (2.60~GHz Intel\textsuperscript{\tiny\textregistered} Xeon\textsuperscript{\tiny\textregistered} E5-2630 v2). Due to the involvement of the advanced nucleus-nucleus collision model G4QMD, the calculations with $^4$He, $^{12}$C and $^{16}$O were slower, delivering up to $6\times 10^6$ histories in each run. However, because of a larger number of secondary particles produced by less diverging ion beams, the statistical uncertainties of dose calculation in phantom voxels were comparable to those of protons.   

Independently of the Monte Carlo modeling with Geant4, 3D histograms with the same parameters were booked and filled for the minibeam arrays on the basis of the DGR approximations of dose from single minibeams,  Sec.~\ref{sec:single_minibeams}. Individual minibeams were placed according to the above-described grid patterns exactly as in the MC modeling. Finally, the distributions of dose from the superpositions of 16 minibeams were obtained exclusively by means of the DGR approximations and were stored as 3D histograms with the same bin size as in MC modeling.

\section{Calculations of PVDR for dose profiles of minibeam arrays}\label{sec:PVDR16}

The peak-to-valley dose ratio (PVDR)~\cite{Prezado2013} is widely considered to be one of the most important  metrics of spatially fractionated dose fields delivered by arrays of minibeams. The calculation of the PVDR values based on approximations is challenging because it requires an accurate description of the dose not only at the peaks, but also at the low-dose valleys, far from the beam axis. 

The dose distributions from arrays of 16 minibeams covering the transverse area of $8\times 8$~mm were considered in the present study, as described in Sec.~\ref{sec:minibeam_arrangements}. In order to mitigate boundary effects and to scale the results to much larger spatially fractionated dose fields,  the PVDR, $D_{\mathrm{PMC}}/D_{\mathrm{VMC}}$, for the dose distributions from MC, and those from the fit, $D_{\mathrm{PFit}}/D_{\mathrm{VFit}}$, were calculated from the doses taken as averages of the doses at certain points in the central inner part of the minibeam array as shown in Fig.~\ref{fig:beam_intensity_transverse}.  Namely, the peak doses at a given depth, $D_{\mathrm{PMC}}$ and $D_{\mathrm{PFit}}$, were obtained as the average of 4 central points for both the rectangular and hexagonal grids, while the valley doses, $D_{\mathrm{VMC}}$ and $D_{\mathrm{VFit}}$ were calculated as the average of 9 or 12 points within the scoring volume at each phantom depth for the rectangular and hexagonal grids, respectively, see Fig.~\ref{fig:beam_intensity_transverse}. Such a selection of points is similar to that in Ref.~\cite{Charyyev2020} where 7 peak and 6 valley doses were considered for the hexagonal unit cell represented by 7 minibeams. In the present work, the additional modeling of 12 minibeams surrounding the unit cell ensured the accurate calculation of all valley doses within the scoring volume. Since only inner points were considered for calculating PVDRs, the dose values in these points were found almost equal to each other for approximated profiles. Some statistical variations of the valley doses are inevitable in MC, but their averages also converged.  This confirmed the validity of our approach. These algorithms were applied to minibeams of 0.3~mm and 0.5~mm FWHM, and to all types of beam particles.

\begin{figure}[tb!]
    \centering
    \includegraphics[width = 0.95\textwidth]{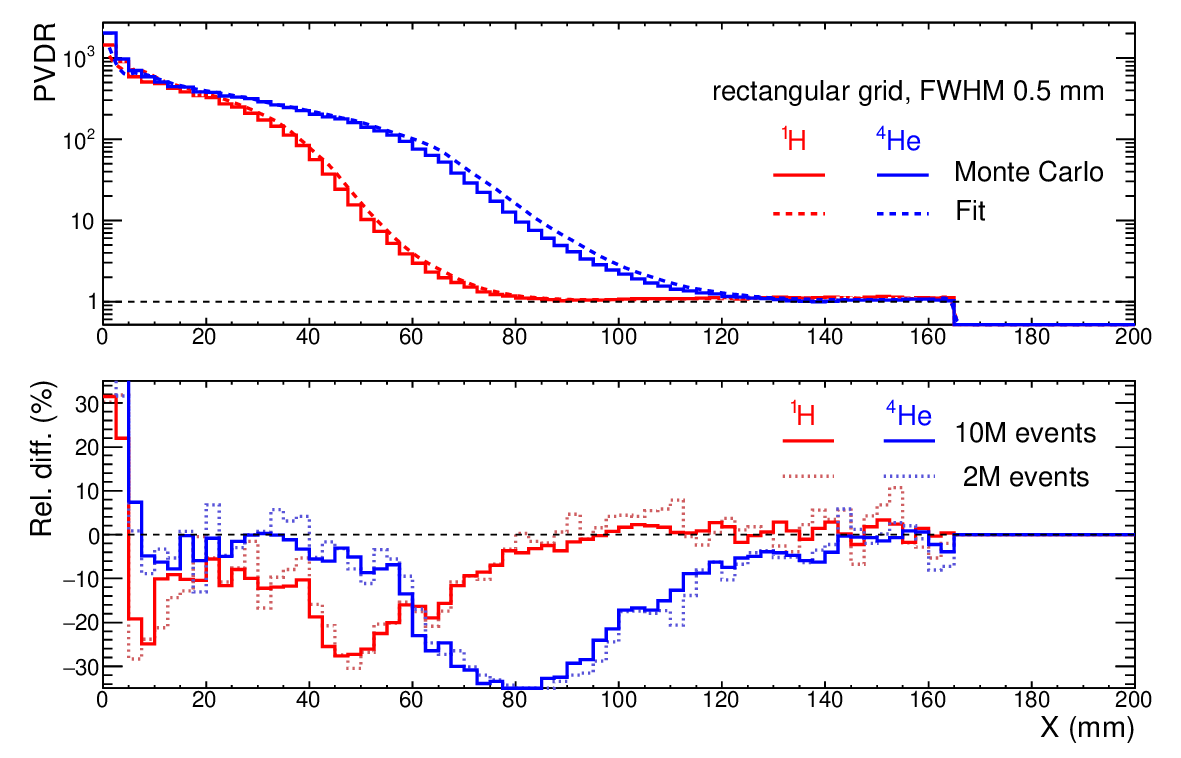}
    \caption{Top panel: peak-to-valley dose ratios calculated for arrays of 16 minibeams of protons and $^{4}$He of 0.5~mm FWHM, placed on the rectangular grid. PVDR calculated for the dose distributions from MC and the fit are presented by solid and dashed lines, respectively. Bottom panel: relative difference between the PVDR values calculated from MC and the fit results. The results for MC simulations with limited numbers of beam particles ($2\times10^{6}$) are shown by doted lines.} 
    \label{fig:PVDR_H1_He4_05mm}
\end{figure}

The PVDR calculated for the dose distributions from MC, and from the fit from arrays of 16 minibeams of protons and $^{4}$He of 0.5~mm FWHM are shown in Fig.~\ref{fig:PVDR_H1_He4_05mm}. The PVDR for protons and $^4$He decrease from $\sim 1000$ to 1 with increasing depth in water to 85~mm and 125~mm, respectively. This evolution from spatially fractionated to more homogeneous dose field demonstrates the main principle of minibeam therapy, although a perfect homogeneity of the target dose is not required. It is only necessary to avoid cold spots within the tumor in order to provide target dose above a prescribed  level~\cite{Sammer2021}.

The PVDRs calculated for the dose  distributions from the MC modeling and those from the fit demonstrate similar  evolution with depth in phantom.  The relative difference  
\begin{equation}
    \Delta = 2\times\frac{D_{\mathrm{PMC}}/D_{\mathrm{VMC}}-D_{\mathrm{PFit}}/D_{\mathrm{VFit}}}{D_{\mathrm{PMC}}/D_{\mathrm{VMC}}+D_{\mathrm{PFit}}/D_{\mathrm{VFit}}}
    \label{eq:Delta}
\end{equation}
was also calculated and presented in the same figure for the quantitative assessment of the similarity between the PVDR values obtained from MC and the fit.

As can be seen, the moduli of the calculated relative difference $\Delta$ are less than 30\% from the entrance to the phantom to the position of the Bragg peak, despite the three orders of magnitude difference between the PVDR values at these two points. This difference includes (1) statistical fluctuations of the dose in individual voxels in the MC modeling; (2) systematic differences between the dose values from the MC and the fit.  Assuming the largest statistical fluctuations (uncertainties) of the peak and valley dose values at the level of 10\%, Sec.~\ref{sec:single_minibeams}, and propagating them to the uncertainty of $\Delta$ given by Eq.~(\ref{eq:Delta}), a value of 27.9\% can be obtained, in full agreement with the magnitude of the fluctuations seen in Fig.~\ref{fig:PVDR_H1_He4_05mm}. As can be seen from this estimate, a good quality of the fit of the dose distribution from a single minibeam is crucial for the accurate calculation of the PVDRs for the minibeam arrays. The convergence of the results of Monte Carlo modeling with respect to the numbers of simulated events is also demonstrated in Fig.~\ref{fig:PVDR_H1_He4_05mm}. The increase of the numbers of events by a factor of 5 reduces the statistical uncertainties of the results, with a similar level of agreement between MC and fit.

\begin{figure}[tbp!]
    \centering
    \includegraphics[width = 0.95\textwidth]{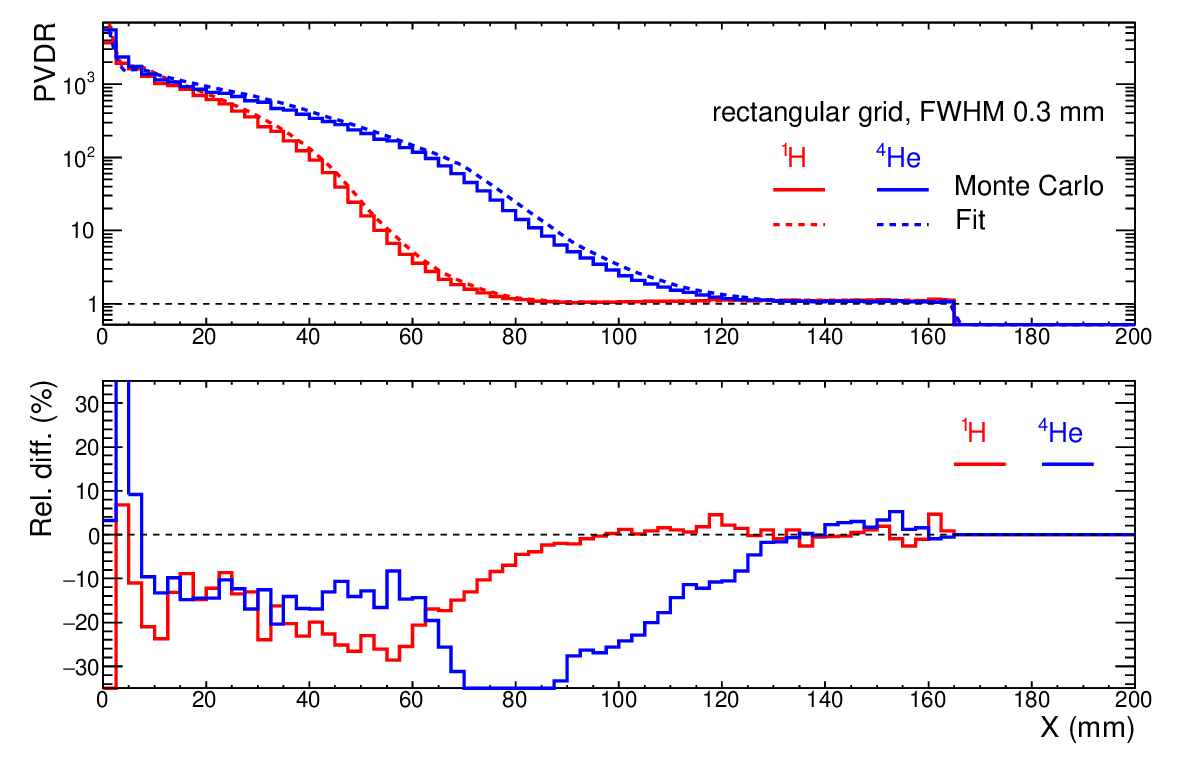}
    \caption{Same as in Fig.~\protect\ref{fig:PVDR_H1_He4_05mm}, but for proton and $^{4}$He minibeams of 0.3~mm FWHM.} 
    \label{fig:PVDR_H1_He4_03mm}
\end{figure}

The depth dependence of the PVDR calculated for the dose distributions from MC and the fit, from arrays of 16 minibeams of protons and $^{4}$He of 0.5~mm FWHM, but arranged on the hexagonal grid is very similar to that for the rectangular grid, and it is not presented here.  Instead, the results for narrower proton and $^{4}$He minibeams of 0.3~mm FWHM arranged on the rectangualr grid are presented in Fig.~\ref{fig:PVDR_H1_He4_03mm}. This combination of narrower beams demonstrates higher PVDR, up to 5000, at the entrance to the phantom. The homogeneous field for protons and $^{4}$He is obtained starting from the same depth of 85~mm and 125~mm, respectively, as with arrays of minibeams of 0.5~mm FWHM. Due to the lower dose values obtained in the valleys for minibeams of 0.3~mm FWHM compared to minibeams of 0.5~mm FWHM, the relative difference between MC and the fit, Fig.~\ref{fig:PVDR_H1_He4_03mm}, is slightly larger in general. 
The results for minibeams of 0.3~mm FWHM arranged on the hexagonal grid are very similar to those of 0.3~mm FWHM on the rectangular grid, including the values of $\Delta$, and they are not presented here. 

\begin{figure}[tbp!]
    \centering
    \includegraphics[width = 0.95\textwidth]{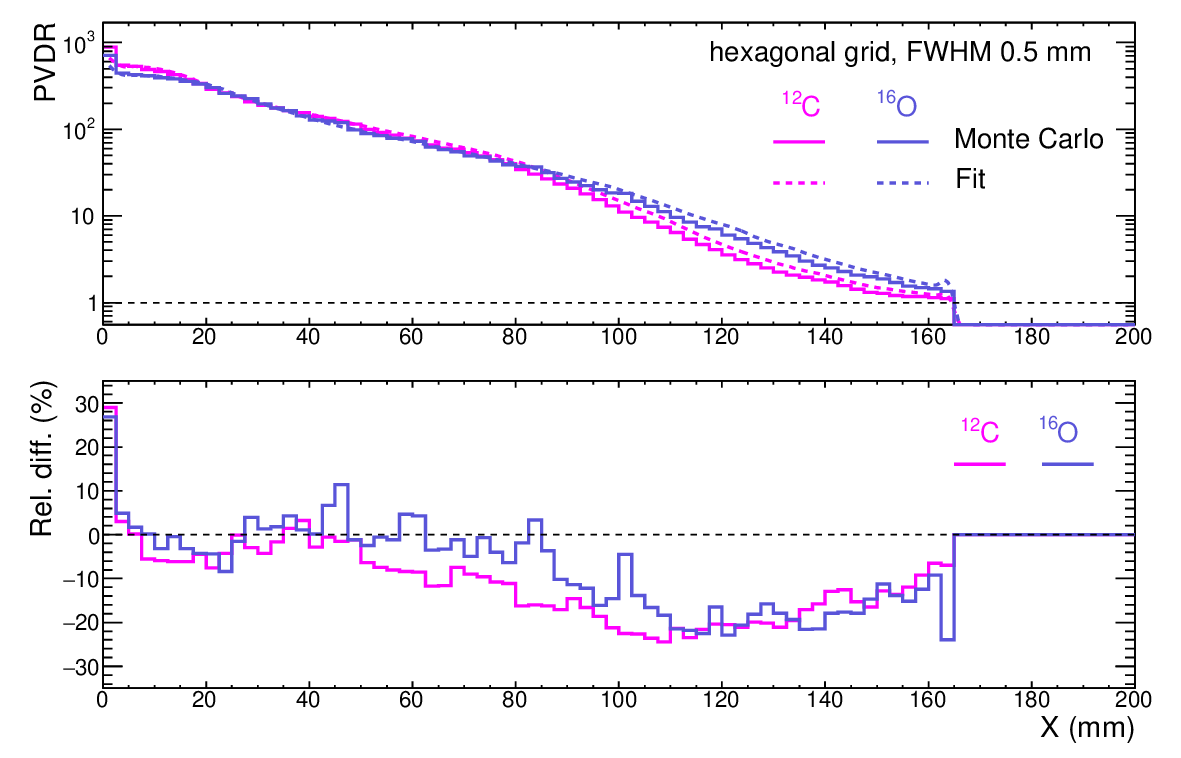}
    \caption{Same as in Fig.~\protect\ref{fig:PVDR_H1_He4_05mm}, but for $^{12}$C and $^{16}$O minibeams of 0.5~mm FWHM arranged on the hexagonal grid.}
    \label{fig:PVDRhex_C12_O16_05mm}
\end{figure}

\begin{figure}[tb!]
    \centering
    \includegraphics[width = 0.95\textwidth]{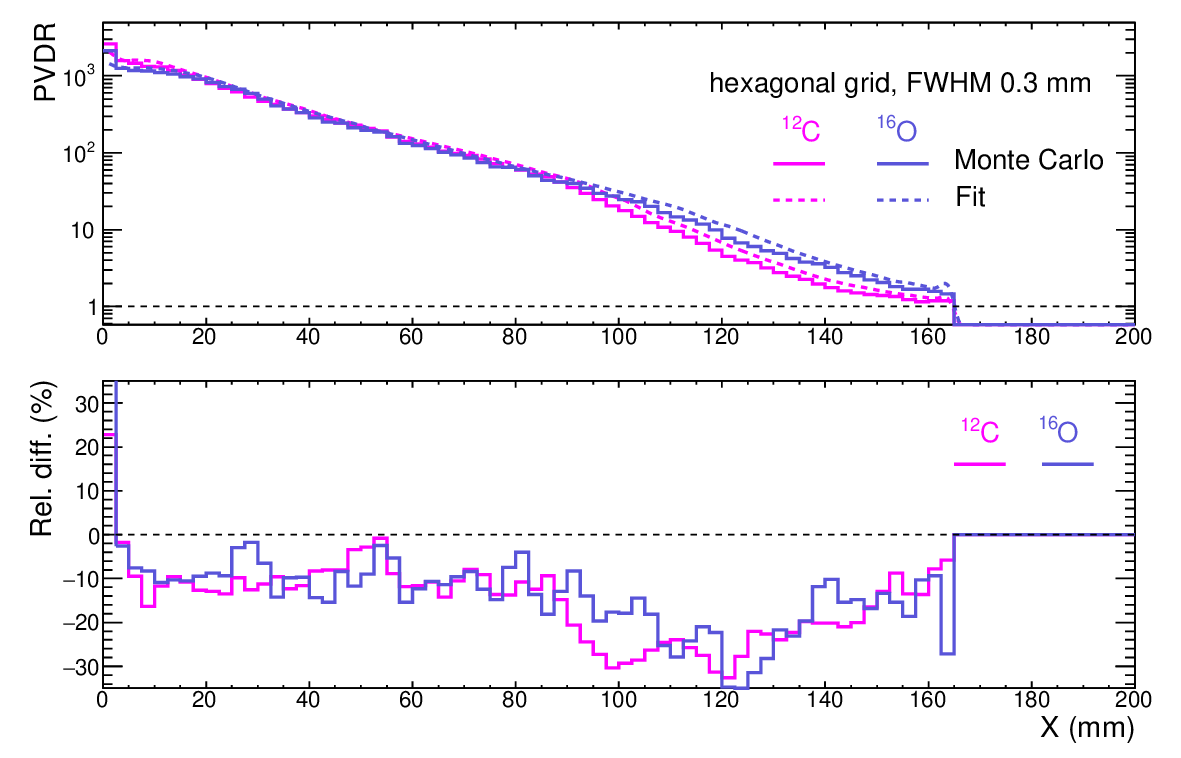}
    \caption{Same as in Fig.~\protect\ref{fig:PVDR_H1_He4_05mm}, but for $^{12}$C and $^{16}$O minibeams of 0.3~mm FWHM arranged on the hexagonal grid.}
    \label{fig:PVDRhex_C12_O16_03mm}
\end{figure}

One can now turn to the results obtained for the minibeams of the heavy projectiles, $^{12}$C and $^{16}$O, see Figs.~\ref{fig:PVDRhex_C12_O16_05mm} and~\ref{fig:PVDRhex_C12_O16_03mm}. In contrast to the arrays of proton and $^{4}$He minibeams, the dose fields of $^{12}$C and $^{16}$O minibeams become homogeneous only close to the Bragg peak due to the reduced lateral deflection of beam particles compared to protons and $^{4}$He.

In view of this, one can give a  preference to the arrangement of minibeams following the hexagonal grid with its denser packing of beam spots, as shown in Fig.~\ref{fig:beam_intensity_transverse}. Therefore, the results for the hexagonal array of 0.5~mm and 0.3~mm FWHM minibeams of $^{12}$C and $^{16}$O are shown in Figs.~\ref{fig:PVDRhex_C12_O16_05mm} and~\ref{fig:PVDRhex_C12_O16_03mm}, respectively. The values of PVDR start with $\sim 1000$ at the phantom entrance and slowly approach to 1 near the Bragg peak. The difference between $^{12}$C and $^{16}$O becomes visible only deeper than 90~mm of water. This general trend of the PVDR obtained from MC is accurately reproduced by the PVDR obtained from the fit. The moduli of the relative difference $\Delta$ between MC and the fit are always less than 30\%, from the entrance of the phantom to the position of the Bragg peak. The magnitude of the $\Delta$ fluctuations is slightly smaller for 0.5~mm FWHM minibeams compared to 0.3~mm FWHM minibeams.

\section{Calculations of DVH for dose profiles of minibeam arrays}\label{sec:DVH16}

Dose-volume histograms (DVHs, also known as cumulative DVHs) are routinely used in radiotherapy treatment planning to verify that the dose is at an appropriate level and is sufficiently uniform throughout the target volume~\cite{Drzymala1991}. DVHs are also used to identify high dose points in adjacent healthy tissues, to evaluate a treatment plan in general and to compare it with another plan. Each bin of a DVH represents the percentage of the volume (Y-axis) that receives the dose greater than or equal to a given dose (X-axis).

Following Ref.~\cite{Sammer2022}, in this work DVH is considered as a possible metrics, complementary to PVDR, for evaluating the degree of spatial dose fractionation at the beginning of the dose plateau (0--20~mm in depth) from the minibeams arranged on the rectangular or hexagonal grid. The degree of dose uniformity can also be accessed by considering the DVH filled in for a target volume (156--160~mm) near the Bragg peak.
In both cases, a central subvolume with transverse dimensions of $4.6\times4.6$~mm$^2$ was used in the calculation of DVHs to avoid boundary effects. The voxel size in calculating DVHs was taken as $0.1\times0.1\times0.1$~mm$^3$. The results obtained for this subvolume as a unit cell can be extended to wider dose fields with the same arrangement of minibeams based on periodic boundary conditions. 

\begin{figure}[htb!]
    \centering
        \includegraphics[width = 0.95\textwidth]{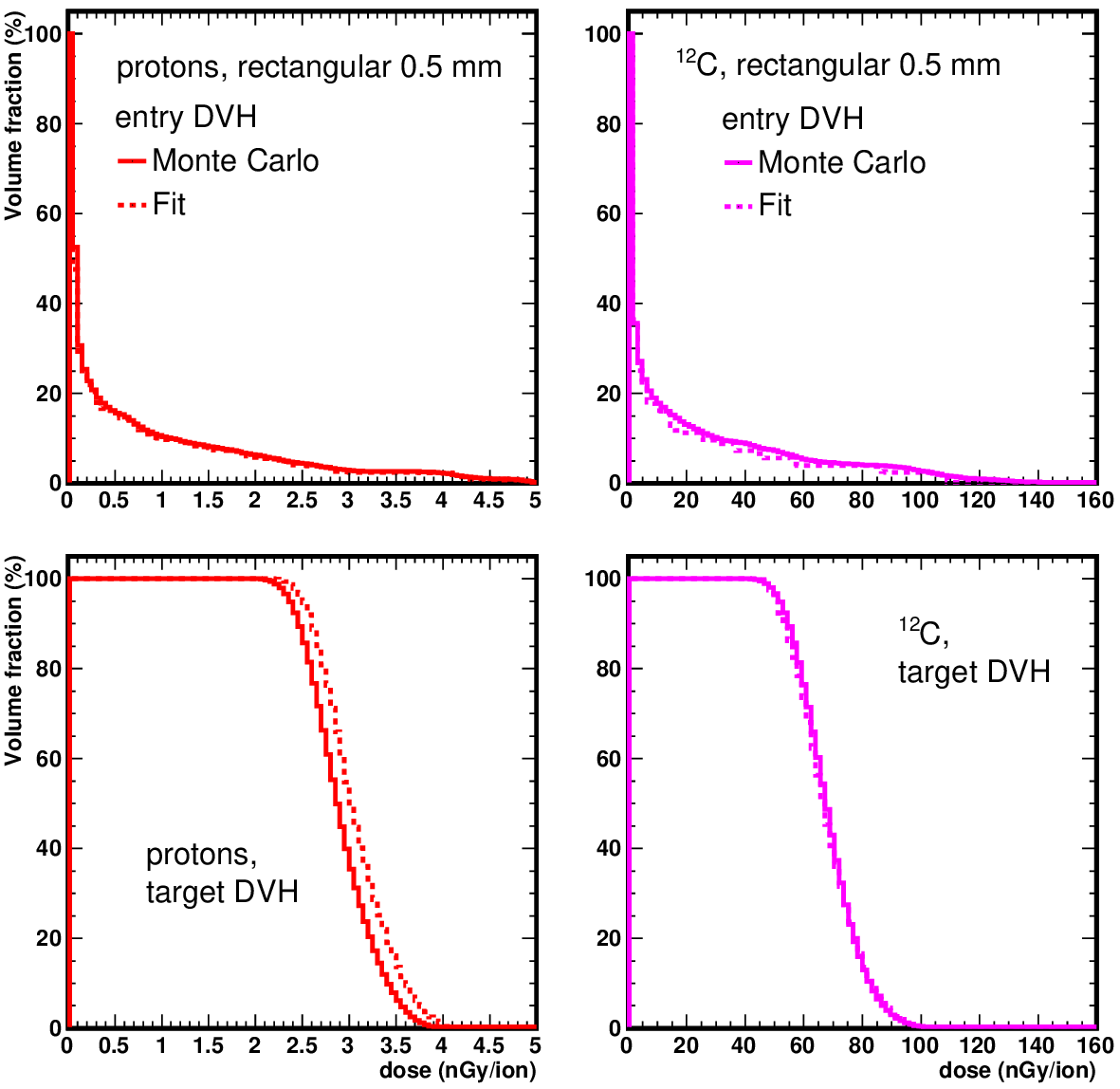} 
    \caption{Dose-volume histograms for arrays of 152~MeV proton and $290A$~MeV $^{12}$C minibeams calculated at the entrance to the phantom (0--20~mm depth) and  the Bragg peak (156--160~mm), top and bottom rows, respectively. Minibeams of 0.5~mm FWHM were arranged on the rectangular grid.  The DVHs calculated from the dose distributions obtained in Monte Carlo modeling and on the basis of the DGR approximations are represented by solid-line and dashed-line histograms, respectively.}
    \label{fig:H1_C12_05mm_DVH}
\end{figure}

Dose-volume histograms for arrays of 152~MeV proton and $290A$~MeV $^{12}$C minibeams calculated at the phantom entrance and the Bragg peak are shown in Fig.~\ref{fig:H1_C12_05mm_DVH} for minibeams of 0.5~mm FWHM arranged on the rectangular grid. The DVHs calculated from the Monte Carlo dose distributions and the DGR approximations are identical for protons and $^{12}$C at the phantom entrance, but small differences are seen in the DVHs calculated for the target volume for protons. In the latter case the DVHs calculated for the dose distributions constructed from the approximated dose profiles show $\sim 7$\% higher dose thresholds delivered to 100\% of the target volume compared to the DVHs obtained from MC. This difference is translated from the 2--10\% difference between the dose values obtained from the fit and MC, see Sec.~\ref{sec:single_minibeams}.  
As can be seen from the DVHs calculated for the entrance subvolume, Fig.~\ref{fig:H1_C12_05mm_DVH}, the doses above 2~nGy/ion and 40~nGy/ion from protons and $^{12}$C, respectively, are delivered to only $\sim 10$\% of this volume. This is a favorable condition to spare healthy tissues~\cite{Zlobinskaya2013,Girst2016}, and it is in contrast to the target volume where the same or higher dose values are delivered to 100\% of the target volume to provide the required control of the tumor. 
\begin{figure}[htb!]
    \centering
        \includegraphics[width = 0.95\textwidth]{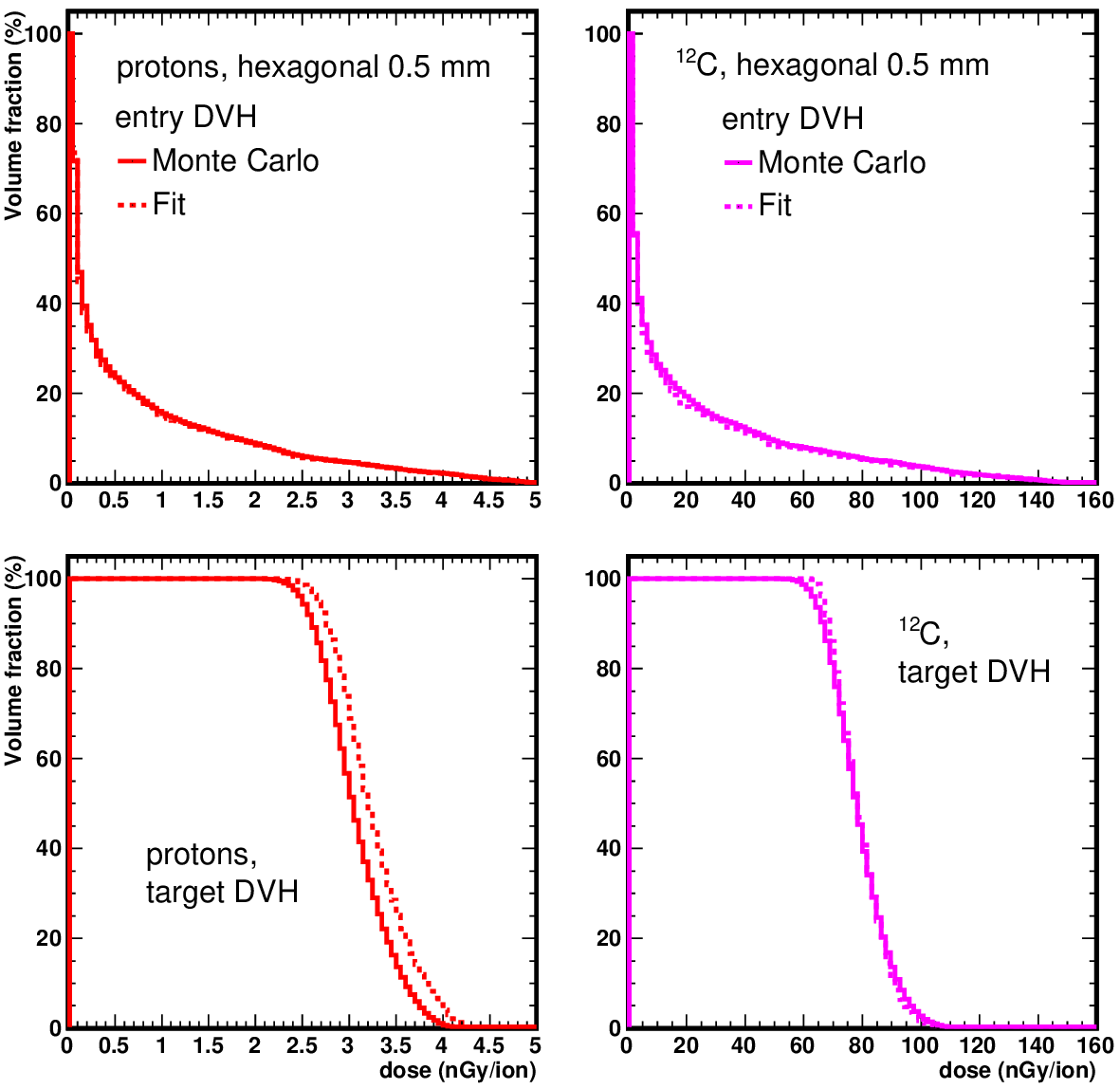}
    \caption{Same as in Fig.~\ref{fig:H1_C12_05mm_DVH}, but for minibeams arranged on the hexagonal grid.}
    \label{fig:H1_C12_05mm_DVHhex}
\end{figure}

The DVHs for arrays of proton and $^{12}$C minibeams of 0.5~mm FWHM were calculated also for the hexagonal grid, see Fig.~\ref{fig:H1_C12_05mm_DVHhex}. With this grid pattern, minibeams are more densely packed in the transverse plane compared to the rectangular grid. As a result, doses higher than 2.3~nGy/ion and 60~nGy/ion are delivered to 100\% of the target volume. As expected, these threshold dose values in the target volume are higher than in the case of the rectangular grid. The DVHs for arrays of $^4$He and $^{16}$O minibeams of 0.3~mm and 0.5~mm were also calculated for the rectangular and hexagonal grids based on MC and the fit, and a good agreement between the two cases was found. This demonstrates the reliability of the DGR approximations obtained in this work for calculating DVHs for arrays of minibeams.

\section{Conclusion} \label{sec:conc}

The distributions of dose from single circular minibeams of protons, $^4$He, $^{12}$C and $^{16}$O of 0.3~mm and 0.5~mm FWHM propagating in water were calculated with Geant4. It was found that for the light projectiles, protons and $^4$He, the expected increase of the local dose near the Bragg peak is completely washed out by the lateral deflection of beam particles from the beam center due to multiple Coulomb scattering in water. As a result, the dose values from protons and $^4$He on the minibeam axis at the phantom entrance are significantly larger than the corresponding values in the Bragg peak region. The same, but less pronounced effect is seen for the dose distributions from minibeams of heavy projectiles, $^{12}$C and $^{16}$O. 

The dose distribution calculated for the 5~mm FWHM pencil-like proton beam demonstrated drastic differences in shape from the 0.5~mm FWHM minibeam. This motivated our search for an alternative to the approximations~\cite{Bellinzona2015,Schwaab2011,Parodi2013} used for pencil-like beams. 

In the present work approximated dose distributions of single mini\-beams of protons, $^{4}$He, $^{12}$C, and $^{16}$O in water were used to build arrays of mini\-beams as a potential alternative to conventional pencil-like beams for a better sparing of healthy tissues. The double-Gauss (DG) and double-Gauss-Rutherford (DGR) functions were considered for approximating the lateral dose profiles of individual minibeams  calculated by Monte Carlo modeling with Geant4. The DGR functions provided a better fit quality compared to the DG functions outside the core formed by primary beam particles.  Upon request to the corresponding author of this paper, MC modeling of individual minibeams with the required FWHM, energy, and projectile type can be performed on demand, with subsequent fitting of the MC dose profiles to provide the parameters of the approximating functions for each case.    

For the considered minibeam parameters the approximation accuracy was accessed by comparing the fitted distributions to the original dose distributions obtained by MC modeling.
The moduli of the relative difference between the dose distributions obtained from the MC modeling and the fit were mostly below 3\% and 6\% for light (protons and $^4$He) and heavy ($^{12}$C and $^{16}$O) projectiles, respectively. Despite the variation of dose between the beam axis and periphery within the three orders of magnitude, this difference reached 10\% only in a very limited number of points in the water phantom.

The dose fields for arrays of 16 parallel circular minibeams in water were constructed using the DGR approximations of single minibeams. The peak-to-valley dose ratios (PVDR) were calculated  for the arrays of minibeams basing directly on the MC modeling as well as on the DGR approximations. Naturally, the differences between the results of MC and fit for single minibeams beams were translated to the differences between the values of PVDR calculated for the dose distributions from the MC and fit. The average moduli of the relative difference were found to be below 30\% for the entire range of the evolution of PVDRs, from 5000 to 1, between the entrance to the phantom and the Bragg peak depth.  Finally, the close similarity of the dose-volume histograms (DVHs) obtained for the distributions of dose from the minibeam arrays from the MC and fit was demonstrated. 

The optimization of the geometry of minibeam arrays is beyond the scope of our work, and it is left for further studies to be based on the developed approximations. In the present work it is demonstrated that these approximations can be used in calculating PVDRs and DVHs for some typical reasonable arrangements of parallel unidirectional minibeam arrays. 

One can conclude that a good agreement between the PVDR values and DVHs obtained for the dose from arrays of parallel minibeam by means of the direct MC modeling and on the basis of approximations justifies the use of the fast calculations based on the DGR approximations instead of the time-consuming Monte Carlo modeling. This means that in future treatment planning systems individual pencil-like beams of a few millimeters FWHM can be replaced by arrays of minibeams without significant additional computational costs. However, it will be necessary to confirm the validity of the proposed approximation for calculations with inhomogeneous phantoms. In future studies the DGR approximations proposed in the present work can be also tested in dose calculations from interlacing minibeams considered in Ref.~\cite{Sammer2021a}. Further studies with accounting for the relative biological efficiency of minibeams of $^4$He, $^{12}$C and $^{16}$O are necessary to extend the present approach to calculations of biological dose and cell survival after minibeam irradiation.

\section*{Acknowledgments}

This study was funded by the Russian Science Foundation (RSF) grant No. 23-25-00285 "Modeling of the physical and biological properties of therapeutic minibeams of protons and nuclei". 
The authors are grateful to RSF for the support.


\bibliographystyle{elsarticle-num} 

\bibliography{Savenkov_minibeams_fit.bib}

\begin{thebibliography}{10}
\expandafter\ifx\csname url\endcsname\relax
  \def\url#1{\texttt{#1}}\fi
\expandafter\ifx\csname urlprefix\endcsname\relax\def\urlprefix{URL }\fi
\expandafter\ifx\csname href\endcsname\relax
  \def\href#1#2{#2} \def\path#1{#1}\fi

\bibitem{Mohan2022}
R.~Mohan, {A review of proton therapy – Current status and future
  directions}, Precis. Radiat. Oncol. 6 (2022) 164--176.
\newblock \href {https://doi.org/10.1002/pro6.1149}
  {\path{doi:10.1002/pro6.1149}}.

\bibitem{Yan2023}
S.~Yan, T.~A. Ngoma, W.~Ngwa, T.~R. Bortfeld, {Global democratisation of proton
  radiotherapy}, Lancet Oncol. 24 (2023) e245--e254.
\newblock \href {https://doi.org/10.1016/S1470-2045(23)00184-5}
  {\path{doi:10.1016/S1470-2045(23)00184-5}}.

\bibitem{Mazal2020}
A.~Mazal, Y.~Prezado, C.~Ares, L.~de~Marzi, A.~Patriarca, R.~Miralbell,
  V.~Favaudon, {FLASH and minibeams in radiation therapy: the effect of
  microstructures on time and space and their potential application to
  protontherapy}, Br. J. Radiol. 93 (2020) 20190807.
\newblock \href {https://doi.org/10.1259/bjr.20190807}
  {\path{doi:10.1259/bjr.20190807}}.

\bibitem{Rudigkeit2024}
S.~Rudigkeit, T.~E. Schmid, A.~C. Dombrowsky, J.~Stolz, S.~Bartzsch, C.-B.
  Chen, N.~Matejka, M.~Sammer, A.~Bergmaier, G.~Dollinger, J.~Reindl,
  {Proton-FLASH: effects of ultra-high dose rate irradiation on an in-vivo
  mouse ear model}, Sci. Rep. 14~(1) (2024) 1418.
\newblock \href {https://doi.org/10.1038/s41598-024-51951-6}
  {\path{doi:10.1038/s41598-024-51951-6}}.

\bibitem{Prezado2013}
Y.~Prezado, G.~R. Fois, {Proton-minibeam radiation therapy: A proof of
  concept}, Med. Phys. 40 (2013) 031712.
\newblock \href {https://doi.org/10.1118/1.4791648}
  {\path{doi:10.1118/1.4791648}}.

\bibitem{Zlobinskaya2013}
O.~Zlobinskaya, S.~Girst, C.~Greubel, {et al}, {Reduced side effects by proton
  microchannel radiotherapy: Study in a human skin model}, Radiat. Environ.
  Biophys. 52 (2013) 123--133.
\newblock \href {https://doi.org/10.1007/s00411-012-0450-9}
  {\path{doi:10.1007/s00411-012-0450-9}}.

\bibitem{Meyer2019}
J.~Meyer, J.~Eley, T.~E. Schmid, S.~E. Combs, R.~Dendale, Y.~Prezado,
  {Spatially fractionated proton minibeams}, Br. J. Radiol. 92 (2019) 20180466.
\newblock \href {https://doi.org/10.1259/bjr.20180466}
  {\path{doi:10.1259/bjr.20180466}}.

\bibitem{Sammer2017}
M.~Sammer, C.~Greubel, S.~Girst, G.~Dollinger, {Optimization of beam
  arrangements in proton minibeam radiotherapy by cell survival simulations},
  Med. Phys. 44 (2017) 6096--6104.
\newblock \href {https://doi.org/10.1002/mp.12566}
  {\path{doi:10.1002/mp.12566}}.

\bibitem{Sammer2021a}
M.~Sammer, S.~Girst, G.~Dollinger, {Optimizing proton minibeam radiotherapy by
  interlacing and heterogeneous tumor dose on the basis of calculated
  clonogenic cell survival}, Sci. Rep. 11 (2021) 3533.
\newblock \href {https://doi.org/10.1038/s41598-021-81708-4}
  {\path{doi:10.1038/s41598-021-81708-4}}.

\bibitem{Girst2016}
S.~Girst, C.~Greubel, J.~Reindl, C.~Siebenwirth, O.~Zlobinskaya, D.~W. Walsh,
  K.~Ilicic, M.~Aichler, A.~Walch, J.~J. Wilkens, G.~Multhoff, G.~Dollinger,
  T.~E. Schmid, {Proton Minibeam Radiation Therapy Reduces Side Effects in an
  In Vivo Mouse Ear Model}, Int. J. Radiat. Oncol. 95 (2016) 234--241.
\newblock \href {https://doi.org/10.1016/j.ijrobp.2015.10.020}
  {\path{doi:10.1016/j.ijrobp.2015.10.020}}.

\bibitem{Prezado2017b}
Y.~Prezado, G.~Jouvion, D.~Hardy, A.~Patriarca, C.~Nauraye, J.~Bergs,
  W.~Gonz{\'{a}}lez, C.~Guardiola, M.~Juchaux, D.~Labiod, R.~Dendale,
  L.~Jourdain, C.~Sebrie, F.~Pouzoulet, {Proton minibeam radiation therapy
  spares normal rat brain: Long-Term Clinical, Radiological and
  Histopathological Analysis}, Sci. Rep. 7 (2017) 14403.
\newblock \href {https://doi.org/10.1038/s41598-017-14786-y}
  {\path{doi:10.1038/s41598-017-14786-y}}.

\bibitem{Sammer2021}
M.~Sammer, A.~C. Dombrowsky, J.~Schauer, K.~Oleksenko, S.~Bicher, B.~Schwarz,
  S.~Rudigkeit, N.~Matejka, J.~Reindl, S.~Bartzsch, A.~Blutke, A.~Feuchtinger,
  S.~E. Combs, G.~Dollinger, T.~E. Schmid, {Normal Tissue Response of Combined
  Temporal and Spatial Fractionation in Proton Minibeam Radiation Therapy},
  Int. J. Radiat. Oncol. 109 (2021) 76--83.
\newblock \href {https://doi.org/10.1016/j.ijrobp.2020.08.027}
  {\path{doi:10.1016/j.ijrobp.2020.08.027}}.

\bibitem{Sammer2022}
M.~Sammer, A.~Rousseti, S.~Girst, J.~Reindl, G.~Dollinger, {Longitudinally
  Heterogeneous Tumor Dose Optimizes Proton Broadbeam, Interlaced Minibeam, and
  FLASH Therapy}, Cancers 14 (2022) 5162.
\newblock \href {https://doi.org/10.3390/cancers14205162}
  {\path{doi:10.3390/cancers14205162}}.

\bibitem{Lansonneur2020}
P.~Lansonneur, H.~Mammar, C.~Nauraye, A.~Patriarca, E.~Hierso, R.~Dendale,
  Y.~Prezado, L.~{De Marzi}, {First proton minibeam radiation therapy treatment
  plan evaluation}, Sci. Rep. 10 (2020) 7025.
\newblock \href {https://doi.org/10.1038/s41598-020-63975-9}
  {\path{doi:10.1038/s41598-020-63975-9}}.

\bibitem{Schneider2019a}
T.~Schneider, A.~Patriarca, Y.~Prezado, {Improving the dose distributions in
  minibeam radiation therapy: Helium ions vs protons}, Med. Phys. 46 (2019)
  3640--3648.
\newblock \href {https://doi.org/10.1002/mp.13646}
  {\path{doi:10.1002/mp.13646}}.

\bibitem{Gonzalez2017}
W.~Gonz{\'{a}}lez, C.~Peucelle, Y.~Prezado, {Theoretical dosimetric evaluation
  of carbon and oxygen minibeam radiation therapy}, Med. Phys. 44 (2017)
  1921--1929.
\newblock \href {https://doi.org/10.1002/mp.12175}
  {\path{doi:10.1002/mp.12175}}.

\bibitem{Gonzalez2018}
W.~Gonz{\'{a}}lez, Y.~Prezado, {Spatial fractionation of the dose in heavy ions
  therapy: An optimization study}, Med. Phys. 45 (2018) 2620--2627.
\newblock \href {https://doi.org/10.1002/mp.12902}
  {\path{doi:10.1002/mp.12902}}.

\bibitem{Grevillot2010}
L.~Grevillot, T.~Frisson, N.~Zahra, {et al.}, {Optimization of GEANT4 settings
  for Proton Pencil Beam Scanning simulations using GATE}, Nucl. Inst. Meth. B
  268 (2010) 3295--3305.
\newblock \href {https://doi.org/10.1016/j.nimb.2010.07.011}
  {\path{doi:10.1016/j.nimb.2010.07.011}}.

\bibitem{Grevillot2011}
L.~Grevillot, D.~Bertrand, F.~Dessy, N.~Freud, D.~Sarrut, {A Monte Carlo pencil
  beam scanning model for proton treatment plan simulation using GATE/GEANT4.},
  Phys. Med. Biol. 56 (2011) 5203--19.
\newblock \href {https://doi.org/10.1088/0031-9155/56/16/008}
  {\path{doi:10.1088/0031-9155/56/16/008}}.

\bibitem{Bellinzona2015}
V.~Bellinzona, M.~Ciocca, A.~Embriaco, A.~Fontana, A.~Mairani, M.~Mori,
  K.~Parodi, {On the parametrization of lateral dose profiles in proton
  radiation therapy}, Phys. Medica 31 (2015) 484--492.
\newblock \href {https://doi.org/10.1016/j.ejmp.2015.05.004}
  {\path{doi:10.1016/j.ejmp.2015.05.004}}.

\bibitem{Parodi2012}
K.~Parodi, A.~Mairani, S.~Brons, B.~G. Hasch, F.~Sommerer, J.~Naumann,
  O.~J{\"{a}}kel, T.~Haberer, J.~Debus, {Monte Carlo simulations to support
  start-up and treatment planning of scanned proton and carbon ion therapy at a
  synchrotron-based facility}, Phys. Med. Biol. 57 (2012) 3759--84.
\newblock \href {https://doi.org/10.1088/0031-9155/57/12/3759}
  {\path{doi:10.1088/0031-9155/57/12/3759}}.

\bibitem{Prezado2021}
Y.~Prezado, {Proton minibeam radiation therapy: a promising therapeutic
  approach for radioresistant tumors}, Comptes Rendus. Biol. 344 (2021)
  409--420.
\newblock \href {https://doi.org/10.5802/crbiol.71}
  {\path{doi:10.5802/crbiol.71}}.

\bibitem{Sammer2019}
M.~Sammer, E.~Zahnbrecher, S.~Dobiasch, S.~Girst, C.~Greubel, K.~Ilicic,
  J.~Reindl, B.~Schwarz, C.~Siebenwirth, D.~W.~M. Walsh, S.~E. Combs,
  G.~Dollinger, T.~E. Schmid, {Proton pencil minibeam irradiation of an in-vivo
  mouse ear model spares healthy tissue dependent on beam size}, PLoS One 14
  (2019) e0224873.
\newblock \href {https://doi.org/10.1371/journal.pone.0224873}
  {\path{doi:10.1371/journal.pone.0224873}}.

\bibitem{Sammer2019a}
M.~Sammer, K.~Teiluf, S.~Girst, C.~Greubel, J.~Reindl, K.~Ilicic, D.~W.~M.
  Walsh, M.~Aichler, A.~Walch, S.~E. Combs, J.~J. Wilkens, G.~Dollinger, T.~E.
  Schmid, {Beam size limit for pencil minibeam radiotherapy determined from
  side effects in an in-vivo mouse ear model}, PLoS One 14 (2019) e0221454.
\newblock \href {https://doi.org/10.1371/journal.pone.0221454}
  {\path{doi:10.1371/journal.pone.0221454}}.

\bibitem{Agostinelli2003}
S.~Agostinelli, J.~Allison, K.~Amako, J.~Apostolakis, {et al}, {Geant4—a
  simulation toolkit}, Nucl. Inst. Meth. Sec. A 506 (2003) 250--303.
\newblock \href {https://doi.org/10.1016/S0168-9002(03)01368-8}
  {\path{doi:10.1016/S0168-9002(03)01368-8}}.

\bibitem{Allison2006}
J.~Allison, K.~Amako, J.~Apostolakis, {et al.}, {Geant4 developments and
  applications}, IEEE Trans. Nucl. Sci. 53 (2006) 270--278.
\newblock \href {https://doi.org/10.1109/TNS.2006.869826}
  {\path{doi:10.1109/TNS.2006.869826}}.

\bibitem{Allison2016}
J.~Allison, K.~Amako, J.~Apostolakis, {et al}, {Recent developments in Geant4},
  Nucl. Inst. Meth. Sec. A 835 (2016) 186--225.
\newblock \href {https://doi.org/https://doi.org/10.1016/j.nima.2016.06.125}
  {\path{doi:https://doi.org/10.1016/j.nima.2016.06.125}}.

\bibitem{Dewey2017}
S.~Dewey, L.~Burigo, I.~Pshenichnov, {et al}, {Lateral variations of
  radiobiological properties of therapeutic fields of $^{1}$H, $^{4}$He,
  $^{12}$C and $^{16}$O ions studied with Geant4 and microdosimetric kinetic
  model}, Phys. Med. Biol. 62 (2017) 5884--5907.
\newblock \href {https://doi.org/10.1088/1361-6560/aa75b2}
  {\path{doi:10.1088/1361-6560/aa75b2}}.

\bibitem{Sato2022}
Y.-h. Sato, D.~Sakata, D.~Bolst, E.~C. Simpson, S.~Guatelli, A.~Haga,
  {Development of a more accurate Geant4 quantum molecular dynamics model for
  hadron therapy}, Phys. Med. Biol. 67 (2022) 225001.
\newblock \href {https://doi.org/10.1088/1361-6560/ac9a9a}
  {\path{doi:10.1088/1361-6560/ac9a9a}}.

\bibitem{Schwaab2011}
J.~Schwaab, S.~Brons, J.~Fieres, {et al}, {Experimental characterization of
  lateral profiles of scanned proton and carbon ion pencil beams for improved
  beam models in ion therapy treatment planning}, Phys. Med. Biol. 56 (2011)
  7813--7827.
\newblock \href {https://doi.org/10.1088/0031-9155/56/24/009}
  {\path{doi:10.1088/0031-9155/56/24/009}}.

\bibitem{Pshenichnov2024}
I.~A. Pshenichnov, U.~A. Dmitrieva, S.~D. Savenkov, A.~O. Svetlichnyi, {Proton
  and Carbon-Ion Minibeam Therapy: From Modeling to Treatment}, Phys. Part.
  Nucl. 55 (2024) 929--934.
\newblock \href {https://doi.org/10.1134/S1063779624700606}
  {\path{doi:10.1134/S1063779624700606}}.

\bibitem{Solie2017}
J.~R. S{\o}lie, H.~E.~S. Pettersen, I.~Meric, {et al.}, {A comparison of proton
  ranges in complex media using GATE/Geant4, MCNP6 and FLUKA} (2017).
\newblock \href {http://arxiv.org/abs/1708.00668} {\path{arXiv:1708.00668}},
  \href {https://doi.org/10.48550/arXiv.1708.00668}
  {\path{doi:10.48550/arXiv.1708.00668}}.

\bibitem{Burigo2013}
L.~Burigo, I.~Pshenichnov, I.~Mishustin, M.~Bleicher, {Microdosimetry of
  radiation field from a therapeutic $^{12}$C beam in water: A study with
  Geant4 toolkit}, Nucl. Inst. Meth. B 310 (2013) 37--53.
\newblock \href {https://doi.org/10.1016/j.nimb.2013.05.021}
  {\path{doi:10.1016/j.nimb.2013.05.021}}.

\bibitem{Regler2000}
M.~Regler, R.~Fr{\"{u}}hwirth, {Efficient modelling of multiple scattering for
  minimum ionizing particles in tissue}, Nucl. Instrum. Methods Phys. Res.,
  Sect. B 170 (2000) 10--20.
\newblock \href {https://doi.org/10.1016/S0168-583X(00)00157-9}
  {\path{doi:10.1016/S0168-583X(00)00157-9}}.

\bibitem{Fruhwirth2000}
R.~Frühwirth, M.~Regler, On the quantitative modelling of core and tails of
  multiple scattering by gaussian mixtures, Nucl. Instrum. Methods Phys. Res.,
  Sect. A 456 (2001) 369--389.
\newblock \href {https://doi.org/10.1016/S0168-9002(00)00589-1}
  {\path{doi:10.1016/S0168-9002(00)00589-1}}.

\bibitem{Parodi2013}
K.~Parodi, A.~Mairani, F.~Sommerer, {Monte Carlo-based parametrization of the
  lateral dose spread for clinical treatment planning of scanned proton and
  carbon ion beams}, J. Radiat. Res. 54~(SUPPL.1) (2013) i91--i96.
\newblock \href {https://doi.org/10.1093/jrr/rrt051}
  {\path{doi:10.1093/jrr/rrt051}}.

\bibitem{Embriaco2017}
A.~Embriaco, E.~V. Bellinzona, A.~Fontana, A.~Rotondi, {On Moli{\`{e}}re and
  Fermi–Eyges scattering theories in hadrontherapy}, Phys. Med. Biol. 62
  (2017) 6290--6303.
\newblock \href {https://doi.org/10.1088/1361-6560/aa7a94}
  {\path{doi:10.1088/1361-6560/aa7a94}}.

\bibitem{Papoulis1968}
A.~Papoulis, {Joint densities with circular symmetry (Corresp.)}, IEEE
  Transactions on Information Theory 14 (1968) 164--165.
\newblock \href {https://doi.org/10.1109/TIT.1968.1054076}
  {\path{doi:10.1109/TIT.1968.1054076}}.

\bibitem{Mayerhofer2021}
M.~Mayerhofer, G.~Datzmann, A.~Degiovanni, V.~Dimov, G.~Dollinger,
  {Magnetically focused 70 MeV proton minibeams for preclinical experiments
  combining a tandem accelerator and a 3 GHz linear post‐accelerator}, Med.
  Phys. 48~(6) (2021) 2733--2749.
\newblock \href {https://doi.org/10.1002/mp.14854}
  {\path{doi:10.1002/mp.14854}}.

\bibitem{Brun1997}
R.~Brun, F.~Rademakers, {ROOT: An object oriented data analysis framework},
  Nucl. Instrum. Meth. A 389 (1997) 81--86.
\newblock \href {https://doi.org/10.1016/S0168-9002(97)00048-X}
  {\path{doi:10.1016/S0168-9002(97)00048-X}}.

\bibitem{BrunSoft}
R.~Brun, F.~Rademakers, P.~Canal, A.~Naumann, O.~Couet, L.~Moneta, V.~Vassilev,
  S.~Linev, D.~Piparo, G.~Ganis, B.~Bellenot, E.~Guiraud, G.~Amadio, P.~Mato,
  M.~Tadel, E.~Tejedor, J.~Blomer, A.~Gheata, S.~Hageboeck, S.~Roiser,
  S.~Wunsch, O.~Shadura, A.~Bose, C.~Cristescu, X.~Valls, R.~Isemann,
  root-project/root: v6.18/02, Tech. rep., CERN (2020).
\newblock \href {https://doi.org/10.5281/zenodo.3895860}
  {\path{doi:10.5281/zenodo.3895860}}.

\bibitem{James1975}
F.~James, M.~Roos, {Minuit: A System for Function Minimization and Analysis of
  the Parameter Errors and Correlations}, Comput. Phys. Commun. 10 (1975)
  343--367.
\newblock \href {https://doi.org/10.1016/0010-4655(75)90039-9}
  {\path{doi:10.1016/0010-4655(75)90039-9}}.

\bibitem{Schneider2022}
T.~Schneider, {Technical aspects of proton minibeam radiation therapy: Minibeam
  generation and delivery}, Phys. Medica 100 (2022) 64--71.
\newblock \href {https://doi.org/10.1016/j.ejmp.2022.06.010}
  {\path{doi:10.1016/j.ejmp.2022.06.010}}.

\bibitem{Charyyev2020}
S.~Charyyev, M.~Artz, G.~Szalkowski, C.~Chang, A.~Stanforth, L.~Lin, R.~Zhang,
  C.~C. Wang, {Optimization of hexagonal-pattern minibeams for spatially
  fractionated radiotherapy using proton beam scanning}, Med. Phys. 47 (2020)
  3485--3495.
\newblock \href {https://doi.org/10.1002/mp.14192}
  {\path{doi:10.1002/mp.14192}}.

\bibitem{Schneider2021a}
T.~Schneider, A.~Patriarca, A.~Degiovanni, M.~Gallas, Y.~Prezado, {Conceptual
  design of a novel nozzle combined with a clinical proton linac for
  magnetically focussed minibeams}, Cancers (Basel). 13 (2021).
\newblock \href {https://doi.org/10.3390/cancers13184657}
  {\path{doi:10.3390/cancers13184657}}.

\bibitem{Datzmann2020}
G.~Datzmann, M.~Sammer, S.~Girst, M.~Mayerhofer, G.~Dollinger, J.~Reindl,
  {Preclinical challenges in proton minibeam radiotherapy: physics and
  biomedical aspects}, Front. Phys. 8 (2020).
\newblock \href {https://doi.org/10.3389/fphy.2020.568206}
  {\path{doi:10.3389/fphy.2020.568206}}.

\bibitem{Drzymala1991}
R.~Drzymala, R.~Mohan, L.~Brewster, {et al.}, {Dose-volume histograms}, Int. J.
  Radiat. Oncol. 21 (1991) 71--78.
\newblock \href {https://doi.org/10.1016/0360-3016(91)90168-4}
  {\path{doi:10.1016/0360-3016(91)90168-4}}.

\end{thebibliography}


\end{document}